\documentclass[twocolumn]{emulateapj}
\usepackage{longtable}
\usepackage{graphicx}
\usepackage{graphics}
\usepackage{color}
\usepackage{epsfig}
\usepackage{rotating}
\usepackage{subfigure}
\usepackage{float}
\usepackage{url}
\usepackage{hyperref}
\usepackage{breakurl}
\usepackage{mathptmx}
\usepackage{amssymb,amsmath}
\usepackage{natbib}

\shortauthors{}
\shorttitle{}

\begin{document}

\title[CHANG-ES VI: Statistical Analysis]{CHANG-ES VI: Probing Supernova Energy Deposition in Spiral Galaxies Through Multi-Wavelength Relationships}

\author{Jiang-Tao Li\altaffilmark{1}, Rainer Beck\altaffilmark{2}, Ralf-J$\rm\ddot{u}$rgen Dettmar\altaffilmark{3}, George Heald\altaffilmark{4}, Judith Irwin\altaffilmark{5}, Megan Johnson\altaffilmark{6}, Amanda A. Kepley\altaffilmark{7}, Marita Krause\altaffilmark{2}, E. J. Murphy\altaffilmark{8}, Elena Orlando\altaffilmark{9}, Richard J. Rand\altaffilmark{10}, A. W. Strong\altaffilmark{11}, Carlos J. Vargas\altaffilmark{12}, Rene Walterbos\altaffilmark{12}, Q. Daniel Wang\altaffilmark{13}, and Theresa Wiegert\altaffilmark{4}} 

\altaffiltext{1}{Department of Astronomy, University of Michigan, 311 West Hall, 1085 S. University Ave, Ann Arbor, MI, 48109-1107, U.S.A.}

\altaffiltext{2}{Max-Planck-Institut f$\rm\ddot{u}$r Radioastronomie, Auf dem H$\rm\ddot{u}$gel 69, 53121, Bonn, Germany}

\altaffiltext{3}{Astronomisches Institut, Ruhr-Universit$\rm\ddot{a}$t Bochum, 44780 Bochum, Germany}

\altaffiltext{4}{Netherlands Institute for Radio Astronomy (ASTRON), Postbus 2, 7990 AA, Dwingeloo, The Netherlands}

\altaffiltext{5}{Department of Physics, Engineering Physics \& Astronomy, Queen’s University, Kingston, ON, Canada, K7L 3N6}

\altaffiltext{6}{CSIRO, PO Box 76, Epping, NSW 1710, Australia}

\altaffiltext{7}{NRAO Charlottesville, 520 Edgemont Road, Charlottesville, VA, 22903-2475, U.S.A.}

\altaffiltext{8}{Observatories of the Carnegie Institution for Science, 813 Santa Barbara Street, Pasadena, CA, 91101, U.S.A.}

\altaffiltext{9}{Physics/Astrophysics Building, 452 Lomita Mall, Stanford University, Stanford, CA, 94305-4085, U.S.A.}

\altaffiltext{10}{Department of Physics and Astronomy, University of New Mexico, 800 Yale Boulevard, NE, Albuquerque, NM, 87131, U.S.A.}

\altaffiltext{11}{Max-Planck-Institut f$\rm\ddot{u}$r extraterrestrische Physik, Garching bei M$\rm\ddot{u}$nchen, Germany}

\altaffiltext{12}{Department of Astronomy, New Mexico State University, PO Box 30001, MSC 4500, Las Cruces, NM, 88003, U.S.A.}

\altaffiltext{13}{Department of Astronomy, University of Massachusetts, 710 North Pleasant St., Amherst, MA, 01003, U.S.A.}



\begin{abstract}
How a galaxy regulates its SNe energy into different interstellar/circumgalactic medium components strongly affects galaxy evolution. Based on the \emph{JVLA} D-configuration C- (6~GHz) and L-band (1.6~GHz) continuum observations, we perform statistical analysis comparing multi-wavelength properties of the CHANG-ES galaxies. The high-quality \emph{JVLA} data and edge-on orientation enable us for the first time to include the halo into the energy budget for a complete radio-flux-limited sample. We find tight correlations of $L_{\rm radio}$ with the mid-IR-based SFR. The normalization of our $I_{\rm 1.6GHz}/{\rm W~Hz^{-1}}-{\rm SFR}$ relation is $\sim$2-3 times of those obtained for face-on galaxies, probably a result of enhanced IR extinction at high inclination. We also find tight correlations between $L_{\rm radio}$ and the SNe energy injection rate $\dot{E}_{\rm SN(Ia+CC)}$, indicating the energy loss via synchrotron radio continuum accounts for $\sim0.1\%$ of $\dot{E}_{\rm SN}$, comparable to the energy contained in CR electrons. The integrated C-to-L-band spectral index is $\alpha\sim0.5-1.1$ for non-AGN galaxies, indicating a dominance by the diffuse synchrotron component. The low-scatter $L_{\rm radio}-{\rm SFR}$/$L_{\rm radio}-\dot{E}_{\rm SN (Ia+CC)}$ relationships have super-linear logarithmic slopes at $\sim2~\sigma$ in L-band ($1.132\pm0.067$/$1.175\pm0.102$) while consistent with linear in C-band ($1.057\pm0.075$/$1.100\pm0.123$). The super-linearity could be naturally reproduced with non-calorimeter models for galaxy disks. Using \emph{Chandra} halo X-ray measurements, we find sub-linear $L_{\rm X}-L_{\rm radio}$ relations. These results indicate that the observed radio halo of a starburst galaxy is close to electron calorimeter, and a galaxy with higher SFR tends to distribute an increased fraction of SNe energy into radio emission (than X-ray).
\end{abstract}

\section{Introduction}\label{sec:Introduction}

The energetic feedback from supernovae (SNe) and massive stellar winds are crucial in setting the basic properties of galaxies, including masses, sizes, and star formation histories, as studied in numerical and semi-analytical models (e.g., \citealt{Somerville14} and references therein). In particular, this stellar feedback is thought to be the most important energy source for the multi-phase circumgalactic medium (CGM) around galaxies without an active galactic nucleus (AGN). The CGM is comprised of multi-phase gas (hot gas $\gtrsim10^6\rm~K$, e.g., \citealt{Li13a}; warm-hot intergalactic medium $\sim10^{5-6}\rm~K$, e.g.,\citealt{Werk13}; cool gas $\sim10^{3-5}\rm~K$, e.g., \citealt{Hoopes99}; and cold gas $<10^3\rm~K$, e.g., \citealt{Heald11}) as well as relativistic particles (cosmic rays; CRs) and magnetic fields (e.g., \citealt{Krause11}), which produce broad-band emission via various emitting mechanisms. Until now, how the galaxies distribute their feedback energy into different interstellar medium (ISM) or CGM phases is still poorly understood (e.g., see \citealt{Wang10,Strong10} and references therein).

Both the diffuse radio and X-ray emission in and around galaxies are thought to be closely related to SNe feedback. The radio continuum emission at GHz frequencies is mainly produced by the synchrotron energy loss of relativistic electrons at $\sim\rm GeV$ energy. These high-energy electrons could be accelerated in supernova remnants (SNRs) or superbubbles (e.g., \citealt{Parizot04,Parizot14}), or from the interaction between the CRs and the ISM (secondary CRs, e.g., \citealt{Strong07}). On the other hand, the soft X-ray emission at $\sim0.5-2\rm~keV$ from galactic halos is thought to be produced by a hot ($kT\gtrsim10^7\rm~K$) tenuous ($n\lesssim10^{-(3-4)}\rm~cm^{-3}$) galactic outflow and its interface with entrained cool cloudlets, instead of the outflow itself (e.g., \citealt{Strickland02,Strickland07,Li09,Li13b,ZhangS14,Li15}). Therefore, the total soft X-ray luminosity is related to both the energy and mass injection rates into the hot phase (e.g., \citealt{Chevalier85,Strickland00,Li11,Zhang14}). As the SN rate and the energy/mass injection rates are all proportional to the star formation rate (SFR) of a galaxy, it is natural that the radio and soft X-ray luminosities of the galactic halos are both correlated to the mid- or far-infrared (IR) luminosity of the galaxy tracing the star formation activity. However, it is still not clear how the diffuse radio and X-ray luminosities of galaxies depend on the SNe feedback energy injection rate (linear dependence or not), and if the radiation efficiency (the fraction of SN energy converted to radiation) in radio and X-ray stays constant at different SFRs or in different types of galaxies (e.g., \citealt{Li13b}). These questions are fundamental for understanding the fate of mass and energy originated from stellar feedback and its role in regulating the structure and evolution of galaxies. 

Tight correlations between radio and IR luminosities of nearby galaxies were discovered as early as the 1980s (e.g., \citealt{Dickey84,deJong85,Helou85}). In most of the studies, the slope of the $L_{\rm radio}-L_{\rm IR}$ relation is found to be close to linear for the global properties of the galaxies (e.g., \citealt{Helou85,Yun01,Vlahakis07}), but the relation shows clear variations at small scales (e.g., $\sim$kpc, \citealt{Murphy06,Murphy08,Heesen14}). \citet{Volk89} developed a calorimeter theory to explain the tight correlation between the non-thermal radio and thermal IR emission, assuming that the source strengths of relativistic electrons and UV photons are both proportional to the SN rate, and the leptonic CRs (electrons and positrons) lose most of their energy within the galactic halo, which is also optically thick to the UV photons heating the dust and producing the IR emission. However, it is widely believed that galaxy disks are not calorimeters even for CR electrons, because many galaxies host extended radio halos (e.g., \citealt{Krause11,Mora13,Irwin13b}). On the other hand, whether or not a galactic halo can be assumed to be a calorimeter for CR electrons is quite uncertain, and could be affected by many factors (e.g., \citealt{Strong10}). \emph{We therefore need to consider the radio halo in the total SNe feedback energy budget.} Slightly but significantly non-linear $L_{\rm radio}-L_{\rm IR}$ relations have also been found and non-calorimeter models have been developed to explain them (e.g., \citealt{Chi90,Price92,Xu94,Niklas97a,Niklas97b,Bell03,Lacki10,Irwin13b,Basu15a}). These studies are extremely important for small scales (e.g., $\sim$kpc) or low-SFR galaxies which do not match the calorimeter conditions.

``Continuum Halos in Nearby Galaxies - an \emph{EVLA} Survey'' (CHANG-ES; \citealt{Irwin12a}) is an unprecedented deep radio continuum survey of a radio-flux-limited sample of 35 nearby edge-on spiral galaxies with the Karl G. Jansky Very Large Array (hereafter \emph{JVLA}) in its commissioning phase. For the first time, the deep \emph{JVLA} observations enable us to detect radio continuum halos in a statistically meaningful sample of nearby galaxies, thus allowing CRs and magnetic fields in the halo to be included in the energy budget of galactic feedback. The major goals of the CHANG-ES project are to investigate the occurrence, origin, and nature of the radio continuum halos, and to probe the disk-halo interaction. A descriptions of the full project is presented in detail in \citet{Irwin12a} (Paper~I). Detailed case studies of three galaxies, NGC~4631, UGC~10288, and NGC~4845, are presented in \citet{Irwin12b} (Paper~II), \citet{Irwin13a} (Paper~III), and \citet{Irwin15} (Paper~V), respectively. 

In Paper~IV (\citealt{Wiegert15}), we presented intensity maps in two frequency bands [C (6~GHz) and L (1.6~GHz)], with different weightings, as well as in-band spectral index maps, polarization maps, and spatially resolved mid-IR measurements of the SFRs. The data products are available to the public in CHANG-ES Data Release 1 at \url{http://www.queensu.ca/changes}. Our early results in Paper~IV indicate that a galaxy with a higher SFR surface density tends to host a larger halo of magnetic fields and CRs. As presented in Paper~IV, the D-configuration measurements in C- and L-bands of the whole sample have little missing flux, therefore they are well-suited for statistical analysis of the total radio continuum luminosity of the galaxies. Furthermore, 16 of the 35 CHANG-ES sample galaxies are also included in the \emph{Chandra} survey of nearby highly-inclined disk galaxies by \citet{Li13a,Li13b,Li14}. The \emph{JVLA} and \emph{Chandra} measurements enable us to directly compare the diffuse radio and X-ray emission from both the galactic disk and halo, which are thought to be related to two of the most energetic phases of galactic feedback: the CR and the hot galactic outflow.

In this paper, we present statistical analysis of the radio continuum and soft X-ray emission of the CHANG-ES galaxies based on the \emph{JVLA} D-configuration and the \emph{Chandra} observations. The paper is organized as follows: In \S\ref{sec:GalPara}, we summarize the data and galaxy parameters used in the present paper and define some subsamples. In \S\ref{sec:StatisticAnalysis}, we present a statistical analysis of the radio continuum emission of the galaxies and compare it to the X-ray and multi-wavelength properties of the galaxies. We further discuss the scientific implications of this statistical analysis in \S\ref{sec:Discussions}. Our major results and conclusions are summarized in \S\ref{sec:Summary}.

\section{Galaxy Parameters and Subsample Definition}\label{sec:GalPara}

\subsection{Galaxy Parameters}\label{subsec:GalPara}

In the present paper, we study the dependence of the diffuse radio continuum luminosity of the whole galaxy (disk+halo, $L_{\rm radio}$) on the multi-wavelength properties of the galaxies, such as the SFR and the SFR surface density, the stellar mass ($M_*$), the total baryon mass ($M_{\rm TF}$), and the energy injection rate from core-collapsed (CC) and Type~Ia SNe ($\dot{E}_{\rm SN,CC}$ and $\dot{E}_{\rm SN,Ia}$). We list these parameters, as well as some other related parameters of the CHANG-ES galaxies, in Table~\ref{Table:GalPara}. More information about the sample galaxies is summarized in \citet{Irwin12a,Wiegert15}. 

\begin{figure*}
\begin{center}
\epsfig{figure=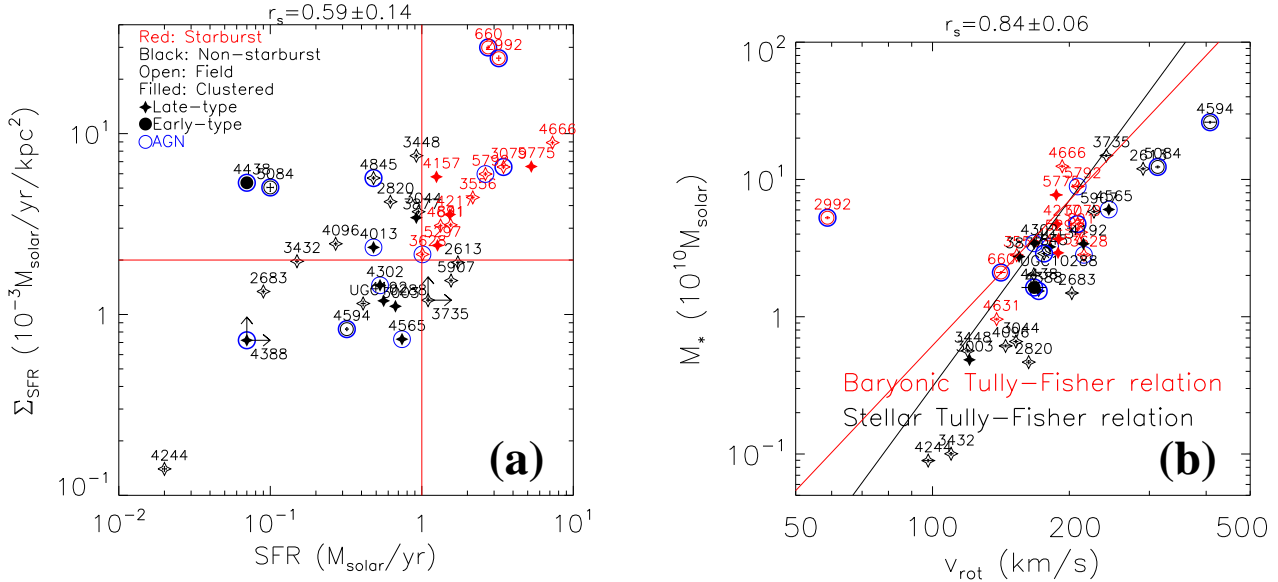,width=1.0\textwidth,angle=0, clip=}
\caption{Basic properties of the CHANG-ES galaxies. (a) The SFR and the SFR surface density ($\Sigma_{\rm SFR}$). Different subsamples are plotted with different symbols: starburst (red) vs. non-starburst (black); field (open) vs. clustered (filled); late-type (diamond) vs. early-type (circle). Symbols for galaxies hosting a radio-bright AGN which could highly affect the radio flux density measurements are surrounded with a thick blue circle, while those with an AGN but unlikely to strongly affect the radio flux measurements are marked with a thin blue circle (``ERB'' and ``RB'' in the last column of Table~\ref{Table:GalPara}, respectively). Data points of NGC~3735 and NGC~4388 are lower limits after removing the AGN \citep{Wiegert15}. The two red lines at ${\rm SFR}=1\rm~M_\odot~yr^{-1}$ and SFR surface density $=2\times10^{-3}\rm~M_\odot~yr^{-1}~kpc^{-2}$ separate starburst and non-starburst galaxies (\S\ref{subsec:SubSample}). The correlation coefficient $r_s$ (\S\ref{sec:StatisticAnalysis}) is calculated for the whole sample. (b) The stellar mass ($M_*$) and gas rotation velocity ($v_{\rm rot}$). $v_{\rm rot}$ is used to derive the baryonic mass of the galaxies adopting the baryonic Tully-Fisher relation from \citet{Bell01} (shown as a red solid line). We also plot the stellar Tully-Fisher relation from \citet{Bell01} for comparison (the black solid line). The correlation coefficient $r_s$ is calculated after removing the interacting galaxy NGC~2992.}\label{Fig:StarburstDefi}
\end{center}
\end{figure*}

The SFR and the SFR surface density ($\Sigma_{\rm SFR}$) are directly quoted from \citet{Wiegert15} (Fig.~\ref{Fig:StarburstDefi}a, but \emph{not} listed in Table~\ref{Table:GalPara}). These star formation parameters are measured with spatially resolved \emph{WISE} 22~$\rm\mu m$ images, after removing bright unresolved nuclear sources in NGC~3735 and NGC~4388, which are assumed to be due to an AGN. The \emph{WISE} 22~$\rm\mu m$ SFRs adopted in Paper~IV and the present paper are systematically lower than those adopted in Paper~I (on average 2.74 times lower). This is partially because in Paper~I we have adopted an old SFR calibrator from \citet{Kennicutt98} which gives a SFR $\sim$17\% higher than the recent result \citep{Kennicutt12}, and partially because the SFR based on total IR luminosity (as adopted in Paper~I) is also systematically higher than that based on mid-IR luminosity (e.g., \emph{WISE} 22~$\rm\mu m$ or \emph{Spitzer} 24~$\rm\mu m$; \citealt{Rieke09,Calzetti10}). In addition, the average total IR to mid-IR luminosity ratio of our sample is $\sim$(18-60)\% higher than those in the literature (e.g.,\citealt{Rieke09,Calzetti10}), which could be an effect of the strong extinction in our highly inclined galaxies. We caution that each star formation tracer has its biases. We herein choose the \emph{WISE} 22~$\rm\mu m$ luminosity as the star formation tracer because it is less affected by the cold dust heated by the general interstellar radiation field which is not related to currect star formation. This cold dust contribution could be important for low-SFR galaxies (e.g., \citealt{Temi09}), which are common in our sample.

We estimate $M_*$ of the galaxies using the \emph{2MASS} K-band magnitude \citep{Skrutskie06} and a color-dependent mass-to-light ratio from \citet{Bell01}. This estimate shows no significant difference from recent estimates of the stellar mass based on \emph{Spitzer} \citep{Zhu10} or \emph{WISE} observations \citep{Jarrett13}. For NGC~3735, which does not have an extinction-corrected B-V color listed in the HyperLeda database \citep{Makarov14}, we assume the mass-to-light ratio to be 1. 

We estimate $M_{\rm TF}$ from the inclination-corrected maximum gas rotation velocity ($v_{\rm rot}$) and the baryonic Tully-Fisher relation (Fig.~\ref{Fig:StarburstDefi}b). $M_{\rm TF}$ includes the contribution from stars and the H{\small~I} gas, but does not include the contribution from molecular gas \citep{Bell01}. Since $M_{\rm TF}$ is derived from the rotation velocity, it provides an alternative distance-independent measurement of the galaxy mass to those derived from the photometry methods (such as $M_*$, which depends on the assumed distance). 

Both the radio continuum and X-ray emission of galaxies are thought to be closely related to SN feedback. However, SNe could be produced by both young and old stellar populations, which are quite different in both temporal and spatial distributions. We therefore need to estimate the energy injection rate from CC and Type~Ia SNe, which are related to young and old stellar populations, respectively, in order to quantitatively combine the contributions from both populations (e.g., \citealt{Li13b}). We estimate $\dot{E}_{\rm SN,CC}$ and $\dot{E}_{\rm SN,Ia}$, adopting the CC SN rate per unit SFR from \citet{Heckman90} and the Type~Ia SN rate per unit stellar mass from \citet{Mannucci05}. We also assume each SN injects a total energy of $10^{51}\rm~ergs$ into the ISM.

\begin{table*}
\vspace{-0.in}
\begin{center}
\small\caption{Parameters of the CHANG-ES Galaxies. Parameters of the CHANG-ES sample galaxies, those marked with $^*$ are included in the \emph{Chandra} sample of nearby edge-on disk galaxies (\citet{Li13a,Li13b,Li14}). $d$ is the distance obtained from \citet{Irwin12a} and further updated in \citet{Wiegert15}. TC is the morphological type code. B-V is the B-V band color corrected for galactic extinction, inclination and redshift. $v_{\rm rot}$ is the maximum gas rotation velocity corrected for inclination. TC, B-V, and $v_{\rm rot}$ are obtained from the HyperLeda database (\citealt{Makarov14}; \url{http://leda.univ-lyon1.fr/}). $M_{\rm TF}$ is the total baryon mass estimated from $v_{\rm rot}$ using the K-band baryonic Tully-Fisher relation \citep{Bell01}. $M_*$ is the stellar mass, estimated from the \emph{2MASS} K-band apparent magnitude \citep{Skrutskie06}, the distance, and the B-V color, adopting a color-dependent stellar mass-to-light ratio \citep{Bell01}. $\dot{E}_{\rm SN,CC}$ and $\dot{E}_{\rm SN,Ia}$ are the core-collapsed and Type~Ia SNe energy injection rate estimated from the SFR (from \citealt{Wiegert15}) and $M_*$, adopting the CC and Ia SN rate from \citet{Heckman90} and \citet{Mannucci05}, respectively. $L_{\rm 6GHz}$ and $L_{\rm 1.6GHz}$ are the C-band and L-band luminosities, estimated by adopting a bandwidth of 2~GHz and 500~MHz in C- and L-bands, respectively. The relatively large error of UGC~10288 is caused by a strong background source \citep{Irwin13a}. $\alpha$ is the integrated spectral index between C- and L-bands of the whole galaxy. ``RF'', ``RB'', ``ERB'' in the ``AGN'' column denote radio faint, radio bright, and extremely radio bright. Only ERB AGN could significantly affect the radio flux measurement.}
\scriptsize
\tabcolsep=3.5pt
\begin{tabular}{lcccccccccccccc}
\hline
Galaxy & $d$ & TC & B-V & $v_{\rm rot}$ & $M_{\rm TF}$ & $M_*$ & $\dot{E}_{\rm SN,CC}$ & $\dot{E}_{\rm SN,Ia}$ & $L_{\rm 6GHz}$ & $L_{\rm 1.6GHz}$ & $\alpha$ & AGN \\
& Mpc & & mag & $\rm km/s$ & $\times10^{10}\rm M_\odot$ & $\times10^{10}\rm M_\odot$ & $\times10^{41}\rm ergs~s^{-1}$ & $\times10^{41}\rm ergs~s^{-1}$ & $\times10^{38}\rm ergs~s^{-1}$ & $\times10^{38}\rm ergs~s^{-1}$ & & \\
\hline
NGC660$^*$ & 12.3 & $1.3\pm1.2$ & 0.71 & 141.4 &  2.08 & $2.10\pm0.03$ & $6.69\pm0.13$ & $0.293\pm0.004$ & $2.382\pm0.048$ & $0.476\pm0.010$ & $-0.17\pm0.02$ & ERB \\
NGC891$^*$ &  9.1 & $3.1\pm0.4$ & 0.70 & 212.1 &  8.63 & $4.13\pm0.06$ & $3.78\pm0.08$ & $0.575\pm0.008$ & $0.414\pm0.014$ & $0.369\pm0.007$ & $0.95\pm0.03$ & RF \\
NGC2613 & 23.4 & $3.1\pm0.5$ & 0.70 & 290.6 & 26.08 & $11.96\pm0.19$ & $4.22\pm0.08$ & $1.67\pm0.03$ & $0.201\pm0.004$ & $0.195\pm0.008$ & $1.02\pm0.03$ & RF \\
NGC2683 &  6.3 & $3.0\pm0.3$ & 0.75 & 202.6 &  7.35 & $1.49\pm0.02$ & $0.220\pm0.004$ & $0.207\pm0.003$ & $0.019\pm0.001$ & $0.016\pm0.002$ & $0.89\pm0.08$ & RF \\
NGC2820 & 26.5 & $5.3\pm0.6$ & 0.23 & 162.8 &  3.41 & $0.467\pm0.013$ & $1.51\pm0.03$ & $0.065\pm0.002$ & $0.321\pm0.007$ & $0.259\pm0.005$ & $0.88\pm0.02$ & RF \\
NGC2992 & 34.0 & $0.9\pm0.5$ & 0.77 &  58.7 &  0.10 & $5.26\pm0.09$ & $7.86\pm0.24$ & $0.734\pm0.012$ & $2.225\pm0.044$ & $1.417\pm0.028$ & $0.70\pm0.02$ & ERB \\
NGC3003 & 25.4 & $4.3\pm0.8$ & 0.28 & 120.6 &  1.19 & $0.485\pm0.010$ & $1.63\pm0.03$ & $0.068\pm0.001$ & $0.167\pm0.008$ & $0.135\pm0.003$ & $0.88\pm0.04$ & RF \\
NGC3044 & 20.3 & $5.5\pm0.7$ & 0.35 & 152.6 &  2.72 & $0.660\pm0.013$ & $2.32\pm0.05$ & $0.092\pm0.002$ & $0.370\pm0.009$ & $0.257\pm0.005$ & $0.76\pm0.02$ & RF \\
NGC3079$^*$ & 20.6 & $6.6\pm0.9$ & 0.53 & 208.4 &  8.11 & $4.73\pm0.07$ & $8.44\pm0.17$ & $0.660\pm0.010$ & $3.711\pm0.074$ & $2.059\pm0.041$ & $0.60\pm0.02$ & ERB \\
NGC3432 &  9.4 & $8.9\pm0.5$ & 0.29 & 109.9 &  0.86 & $0.100\pm0.002$ & $0.366\pm0.007$ & $0.0140\pm0.0003$ & $0.056\pm0.001$ & $0.044\pm0.001$ & $0.86\pm0.02$ & RF \\
NGC3448 & 24.5 & $1.8\pm2.5$ & 0.31 & 119.5 &  1.15 & $0.564\pm0.011$ & $2.24\pm0.04$ & $0.079\pm0.002$ & $0.295\pm0.006$ & $0.165\pm0.003$ & $0.60\pm0.02$ & RF \\
NGC3556$^*$ & 14.1 & $6.0\pm0.3$ & 0.57 & 153.2 &  2.76 & $2.81\pm0.04$ & $5.29\pm0.11$ & $0.392\pm0.006$ & $0.376\pm0.022$ & $0.346\pm0.007$ & $0.97\pm0.05$ & RF \\
NGC3628$^*$ &  8.5 & $3.1\pm0.4$ & 0.68 & 215.4 &  9.11 & $2.83\pm0.04$ & $2.46\pm0.05$ & $0.394\pm0.006$ & $0.319\pm0.006$ & $0.228\pm0.005$ & $0.78\pm0.02$ & RB \\
NGC3735 & 42.0 & $5.3\pm0.6$ & 0.00 & 241.1 & 13.55 & $14.92\pm0.21$ & $2.68\pm0.05$ & $2.08\pm0.03$ & $1.051\pm0.021$ & $0.858\pm0.017$ & $0.88\pm0.02$ & RF \\
NGC3877$^*$ & 17.7 & $5.1\pm0.5$ & 0.65 & 155.1 &  2.87 & $2.74\pm0.04$ & $2.24\pm0.04$ & $0.382\pm0.006$ & $0.097\pm0.002$ & $0.080\pm0.002$ & $0.89\pm0.02$ & RF \\
NGC4013$^*$ & 16.0 & $3.1\pm0.5$ & 0.83 & 181.4 &  4.99 & $3.23\pm0.05$ & $1.17\pm0.02$ & $0.450\pm0.007$ & $0.077\pm0.002$ & $0.058\pm0.001$ & $0.82\pm0.02$ & RB \\
NGC4096 & 10.3 & $5.3\pm0.6$ & 0.50 & 144.8 &  2.26 & $0.613\pm0.010$ & $0.659\pm0.013$ & $0.086\pm0.001$ & $0.042\pm0.001$ & $0.036\pm0.001$ & $0.94\pm0.02$ & RF \\
NGC4157 & 15.6 & $3.3\pm0.7$ & 0.64 & 188.9 &  5.75 & $2.92\pm0.04$ & $3.05\pm0.06$ & $0.407\pm0.006$ & $0.321\pm0.006$ & $0.269\pm0.005$ & $0.90\pm0.02$ & RF \\
NGC4192 & 13.6 & $2.6\pm0.8$ & 0.66 & 214.8 &  9.03 & $3.40\pm0.05$ & $1.37\pm0.03$ & $0.474\pm0.007$ & $0.107\pm0.002$ & $0.089\pm0.002$ & $0.89\pm0.02$ & RF \\
NGC4217$^*$ & 20.6 & $3.1\pm0.4$ & 0.75 & 187.6 &  5.62 & $4.74\pm0.07$ & $3.73\pm0.07$ & $0.661\pm0.010$ & $0.360\pm0.007$ & $0.283\pm0.006$ & $0.86\pm0.02$ & RF \\
NGC4244$^*$ &  4.4 & $5.9\pm0.6$ & 0.41 &  97.8 &  0.57 & $0.090\pm0.002$ & $0.049\pm0.001$ & $0.0125\pm0.0003$ & $0.0042\pm0.0003$ & $0.0021\pm0.0001$ & $0.52\pm0.06$ & RF \\
NGC4302 & 19.4 & $5.4\pm0.7$ & 0.74 & 167.4 &  3.76 & $3.43\pm0.05$ & $1.29\pm0.03$ & $0.478\pm0.007$ & $0.108\pm0.003$ & $0.102\pm0.002$ & $0.99\pm0.02$ & RB \\
NGC4388$^*$ & 16.6 & $2.8\pm0.6$ & 0.57 & 171.2 &  4.07 & $1.54\pm0.02$ & $0.171\pm0.003$ & $0.215\pm0.003$ & $0.410\pm0.008$ & $0.216\pm0.004$ & $0.56\pm0.02$ & ERB \\
NGC4438$^*$ & 10.4 & $0.6\pm1.3$ & 0.76 & 167.3 &  3.76 & $1.63\pm0.02$ & $0.171\pm0.005$ & $0.228\pm0.003$ & $0.141\pm0.003$ & $0.085\pm0.002$ & $0.66\pm0.02$ & ERB \\
NGC4565$^*$ & 11.9 & $3.3\pm0.7$ & 0.68 & 244.3 & 14.18 & $6.04\pm0.08$ & $1.81\pm0.04$ & $0.842\pm0.012$ & $0.143\pm0.004$ & $0.129\pm0.003$ & $0.96\pm0.02$ & RB \\
NGC4594$^*$ & 12.7 & $1.1\pm0.3$ & 0.88 & 408.0 & 85.82 & $26.12\pm0.36$ & $0.781\pm0.016$ & $3.64\pm0.05$ & $0.495\pm0.010$ & $0.090\pm0.002$ & $-0.24\pm0.02$ & ERB \\
NGC4631$^*$ &  7.4 & $6.6\pm0.7$ & 0.39 & 138.4 &  1.93 & $0.960\pm0.015$ & $3.25\pm0.06$ & $0.134\pm0.002$ & $0.373\pm0.010$ & $0.355\pm0.012$ & $1.00\pm0.03$ & RF \\
NGC4666$^*$ & 27.5 & $5.1\pm0.7$ & 0.64 & 192.9 &  6.19 & $12.48\pm0.18$ & $17.79\pm0.36$ & $1.74\pm0.03$ & $2.268\pm0.045$ & $1.830\pm0.037$ & $0.88\pm0.02$ & RF \\
NGC4845 & 17.0 & $2.3\pm0.8$ & 0.78 & 176.0 &  4.48 & $2.89\pm0.05$ & $1.17\pm0.02$ & $0.403\pm0.007$ & $2.981\pm0.059$ & $0.397\pm0.008$ & $-0.47\pm0.02$ & ERB \\
NGC5084 & 23.4 & $-2.0\pm0.4$ & 0.93 & 312.9 & 33.81 & $12.33\pm0.19$ & $0.244\pm0.012$ & $1.72\pm0.03$ & $0.473\pm0.013$ & $0.133\pm0.003$ & $0.09\pm0.03$ & ERB \\
NGC5297 & 40.4 & $4.9\pm0.7$ & 0.50 & 189.5 &  5.81 & $3.69\pm0.07$ & $3.10\pm0.06$ & $0.515\pm0.009$ & $0.262\pm0.016$ & $0.238\pm0.005$ & $0.96\pm0.05$ & RF \\
NGC5775$^*$ & 28.9 & $5.2\pm0.6$ & 0.66 & 187.2 &  5.57 & $7.72\pm0.10$ & $12.88\pm0.26$ & $1.08\pm0.01$ & $1.487\pm0.030$ & $1.274\pm0.025$ & $0.92\pm0.02$ & RF \\
NGC5792 & 31.7 & $3.0\pm0.3$ & 0.69 & 208.6 &  8.15 & $8.89\pm0.16$ & $6.42\pm0.13$ & $1.24\pm0.02$ & $0.486\pm0.010$ & $0.347\pm0.007$ & $0.78\pm0.02$ & RB \\
NGC5907 & 16.8 & $5.3\pm0.6$ & 0.62 & 226.6 & 10.89 & $5.82\pm0.09$ & $3.81\pm0.08$ & $0.812\pm0.012$ & $0.348\pm0.007$ & $0.307\pm0.006$ & $0.94\pm0.02$ & RF \\
UGC10288 & 34.1 & $5.3\pm0.6$ & 0.66 & 167.1 &  3.74 & $2.03\pm0.05$ & $1.00\pm0.03$ & $0.284\pm0.007$ & $0.043\pm0.001$ & $0.031\pm0.003$ & $0.79\pm0.09$ & RF \\
\hline
\end{tabular}\label{Table:GalPara}
\end{center}
\end{table*}

The radio measurements are based on the \emph{JVLA} D-configuration observations presented in \citet{Wiegert15}. The edge-on orientation of the CHANG-ES galaxies highlights the multi-wavelength emission from the galactic halos. Compared to previous observations, our \emph{JVLA} observations are deep enough to detect the extended halo component and have high enough resolution to detect and remove bright point-like sources. Therefore, with the high-sensitivity, broad-band \emph{JVLA} data we are not only able to use the most accurate radio continuum measurements of a complete flux-limited nearby edge-on spiral galaxy sample, but are also able to study the spatial variation of the radio continuum properties in both disk and halo of individual galaxies. Some preliminary results of these spatially-resolved studies are presented in \citet{Wiegert15} and will be further discussed in companion CHANG-ES papers. We have removed all the disconnected point-like sources around the galaxies but have included the AGN and any extended extraplanar features in the final radio flux measurement. Therefore, the radio luminosities quoted in the present paper are the emission from the whole galaxy, including the contributions from both the disk and the halo. We adopt bandwidths of $2\rm~GHz$ and $500\rm~MHz$ in C- and L-bands in order to convert the radio flux density to luminosity ($L_{\rm6GHz}$ and $L_{\rm1.6GHz}$) for the convenience of comparison to $L_{\rm X}$ and $\dot{E}_{\rm SN}$. We also calculate the integrated spectral index between C- and L-bands, which is defined as $S(\nu)\propto\nu^{-\alpha}$, where $S(\nu)$ is the global flux density in C- and L-bands, respectively. $L_{\rm6GHz}$, $L_{\rm1.6GHz}$, and $\alpha$ are also listed in Table~\ref{Table:GalPara}. 

There is an amount of missing flux density that mainly depends on the spatial scale of the emission features and the array configuration. For D-configuration in L-band, the largest angular scale is $\sim16.2^\prime$, larger than the optical diameter ($D_{\rm25}$) of all the CHANG-ES galaxies \citep{Irwin12a}. In most of the cases, even some starburst galaxies, the vertical extension of the radio continuum emission is no larger than the extension in the radial direction (e.g., \citealt{Irwin13b}). Therefore, the L-band data should include all the emission from the galaxy. For D-configuration in C-band, however, the largest angular scale is $\sim4^\prime$, smaller than the optical diameter of most of the CHANG-ES galaxies. We are thus likely resolving out some extended diffuse emission. Comparison with the catalog of \citet{Gregory91} suggests that we are typically missing $\lesssim50\%$ of the diffuse flux density in the more extended CHANG-ES galaxies (assuming a spectral index of $\alpha\sim1.0$ and ignoring the AGN). To accurately measure the amount of missing flux density, deep single-dish observations are required. In general, larger galaxies and/or intense star forming galaxies may have large-scale radio continuum features, so a larger fraction of missing flux density in C-band.

Sixteen of the CHANG-ES galaxies are included in the \emph{Chandra} sample of nearby edge-on disk galaxies \citep{Li13a,Li13b,Li14} (Table~\ref{Table:GalPara}); their soft X-ray luminosities are quoted from \citet{Li13a}. \emph{Chandra} data calibration and analysis are presented in \citet{Li13a} and statistical analysis of the X-ray properties of the galaxies are presented in \citet{Li13b,Wang15}. We further compare the X-ray measurements to cosmological hydrodynamical simulations of hot galactic coronae in \citet{Li14}. In particular, the edge-on galaxies often show clear absorption features at the location of the galactic disk \citep{Li13a}. In order to have a uniform comparison among different galaxies, we first fit the extraplanar diffuse X-ray profile with an exponential model after removing the point-like sources and the regions heavily affected by disk absorption. The X-ray luminosities are measured within a vertical distance of five times the exponential scale height, after quantitatively removing the contributions from unresolved stellar sources and correcting for the filtered absorbing galactic disk by interpolating the diffuse X-ray intensity profile \citep{Li13a}. The X-ray luminosities are scaled to the distances of the CHANG-ES galaxies adopted in the present paper.

\subsection{Subsample Definition}\label{subsec:SubSample}

The presence of radio-bright AGN and sometimes even extended features related to the AGN could significantly affect the radio flux density measurements of the galaxies. The identification and properties of the AGN of the CHANG-ES galaxies will be presented in detail in a separate paper. In the present paper, we identify 13 galaxies with \emph{radio bright} AGN, satisfying at least one of the following criteria:  (1) The galaxy has an unresolved radio core in higher resolution (B- or C-configuration) radio images which also shows polarization features and/or negative radio spectral index $\alpha$; (2) There are lines of evidence for the existence of an outflow (e.g., radio jets or bubbles); (3) By comparing with archival radio data, there is significant radio flux variability of the core of the galaxy; (4) The galaxy has abnormal IR color in \emph{WISE} bands in the nuclear region; (5) An AGN is identified in the literature (see a summary of the archival identification of AGN for the CHANG-ES galaxies in \citealt{Irwin12a}). 

All the galaxies identified as hosting a radio bright AGN satisfy criterion (1). We further define a galaxy with an extremely radio bright (``ERB'') AGN only if at least one of the additional criteria (2) or (3) is satisfied. These eight ``ERB'' galaxies are: NGC~660 (jet, variability; \citealt{Argo15}), NGC~2992 (extended nuclear radio structure, \citealt{Gallimore06}), NGC~3079 (radio lobe, \citealt{Gallimore06}), NGC~4388 (jet, \citealt{Hummel91}), NGC~4438 (jet, \citealt{Hummel91}; to be presented in a separate CHANG-ES paper, Damas et al., in prep.), NGC~4594 (jet, \citealt{Hada13}; variability, \citealt{Krause06}), NGC~4845 (variability, \citealt{Irwin15}), and NGC~5084 (extended nuclear radio structure; the results will be presented in a separate CHANG-ES paper). We identify five other galaxies with radio bright (``RB'') AGN (Table~\ref{Table:GalPara}). Although some of the galaxies labeled with ``ERB'' are indeed classified as radio quiet in the literature (e.g., \citealt{Gallimore06}), in the CHANG-ES sample their radio flux densities are in general significantly affected by the presence of the AGN. In the following sections, we will compare the results of statistical analysis with and without these ``ERB'' galaxies included in the sample, while ``RB'' galaxies are still treated as normal galaxies with radio faint nuclei in all the calculations. All the ``ERB'' and ``RB'' galaxies will be plotted with different symbols in the figures presented in this paper.

We also include the Milky Way (MW) in some of our plots for comparison. The L-band radio luminosity is obtained from \citet{Strong10} and converted to the same frequency and bandwidth as the CHANG-ES sample, assuming a spectral index of $\alpha\sim0.8$. The error is calculated from the standard deviation of the luminosities of different models discussed in \citet{Strong10}. The final L-band luminosity of the Milky Way adopted in this paper is $L_{\rm1.6GHz,MW}=(6.1\pm2.0)\times10^{36}\rm~ergs~s^{-1}$. The SFR of the Milky Way is taken to be ${\rm SFR_{MW}}=(1.065\pm0.385)\rm~M_\odot~yr^{-1}$ \citep{Robitaille10}. The stellar mass of the Milky Way is taken to be ${\rm M_{\rm *,MW}}=(6.43\pm0.63)\times10^{10}\rm~M_\odot$ \citep{McMillan11}. The soft X-ray luminosity of the Milky Way is estimated from the \emph{ROSAT} All Sky Survey: $L_{\rm X,MW}\sim2\times10^{39}\rm~ergs~s^{-1}$ \citep{Snowden97}.

Similar to \citet{Li13a}, we also define some subsamples for the convenience of comparison among different types of galaxies. Different subsamples described below will be plotted as different symbols in the following statistical analysis. 

We use the local galaxy number density ($\rho$; \citealt{Irwin12a}) to characterize the galaxy environment, and define galaxies with $\rho\leq0.6$ as being in the field, while those with $\rho>0.6$ as being clustered. This is just a general classification to characterize the global environment of the galaxies. We caution that some galaxies in poor groups or interacting galaxies with close companions (e.g., NGC~3432, \citealt{English97}) have been classified as in the field, because they are not in rich clusters with a dense intra-cluster medium (ICM). 

We also separate early- and late-type disk galaxies with their morphological type code (TC, Table~\ref{Table:GalPara}). Galaxies with $\rm TC\leq1.5$ are Sa or S0 and are defined as early-type, while others are late-type. This is again a general classification without considering the highly distorted morphology of some interacting galaxies (e.g., NGC~660). 

The star formation properties adopted in the present paper are estimated based on spatially-resolved mid-IR images \citep{Wiegert15}, which allow us to estimate the SFR surface density, rather than just the global integrated IR luminosities adopted in \citet{Li13a}. We therefore adopt a new criterion to define starburst and non-starburst galaxies, i.e., those with both ${\rm SFR}\geq1\rm~M_\odot~yr^{-1}$ and SFR surface density $\geq2\times10^{-3}\rm~M_\odot~yr^{-1}~kpc^{-2}$ are starbursts while others are non-starbursts (Fig.~\ref{Fig:StarburstDefi}a). Such a definition of a starburst galaxy is roughly consistent with the criteria adopted in \citet{Li13a} based on IR luminosity and color. It is expected from both theoretical and observational studies that only starburst galaxies with a SFR surface density above a threshold should have a significant amount of extraplanar hot gas blown out of the disk \citep{Strickland04,Li13b}. NGC~4438 and NGC~5084 have extremely high SFR surface density (Fig.~\ref{Fig:StarburstDefi}a). In these two galaxies, most of the \emph{WISE} IR flux comes from the nucleus, but it is uncertain how much of it is due to star formation or an AGN.

\section{Statistical analysis}\label{sec:StatisticAnalysis}

Similar to \citet{Li13b}, we also utilize the Spearman's rank order correlation coefficient ($r_s$; by definition, $|r_s|<1$, negative means anti-correlation) to describe the tightness of a correlation. We consider $|r_s|\geq0.6$ as a tight correlation, $0.3\leq|r_s|<0.6$ as a weak correlation, and $|r_s|<0.3$ as no correlation. We further characterize some important relations with simple expressions, with the corresponding root mean square ($rms$) around the best-fit relations describing the dispersion. In order to fit a relation and to estimate the uncertainties of $r_s$ and $rms$, we first generate 1000 bootstrap-with-replacement samples of the primary data and then resample each data point, assuming a normal distribution. For each re-generated sample, we fit the data with the same model and measure the model parameters, as well as $r_s$ and $rms$, in the same way. These measurements are then rank-ordered; their 68\% percentiles around the original fitting parameters, $r_s$ and $rms$, are finally used as the estimates of their $1-\sigma$ errors presented in figures and equations. Whenever we need to fit the logarithmic slope of a relation, we perform a linear fit with the IDL function ``\emph{ladfit}'' in a logarithmic scale for both $x$ and $y$ axes. The ``\emph{ladfit}'' function uses a robust least absolute deviation method and has the advantage of being less affected by some significantly outlying data points. In the following sections, the eight galaxies whose radio fluxes are highly affected by the AGN (``ERB'' in Table~\ref{Table:GalPara}) will not be included in quantitative calculations of $r_s$ and $rms$ presented on the figures, but we will compare the fitted relations between different galaxy parameters including these ``ERB'' galaxies or not.

\subsection{Statistical analysis of the diffuse radio continuum emission}\label{subsec:StatisticRadioHalo}

\begin{figure*}
\begin{center}
\epsfig{figure=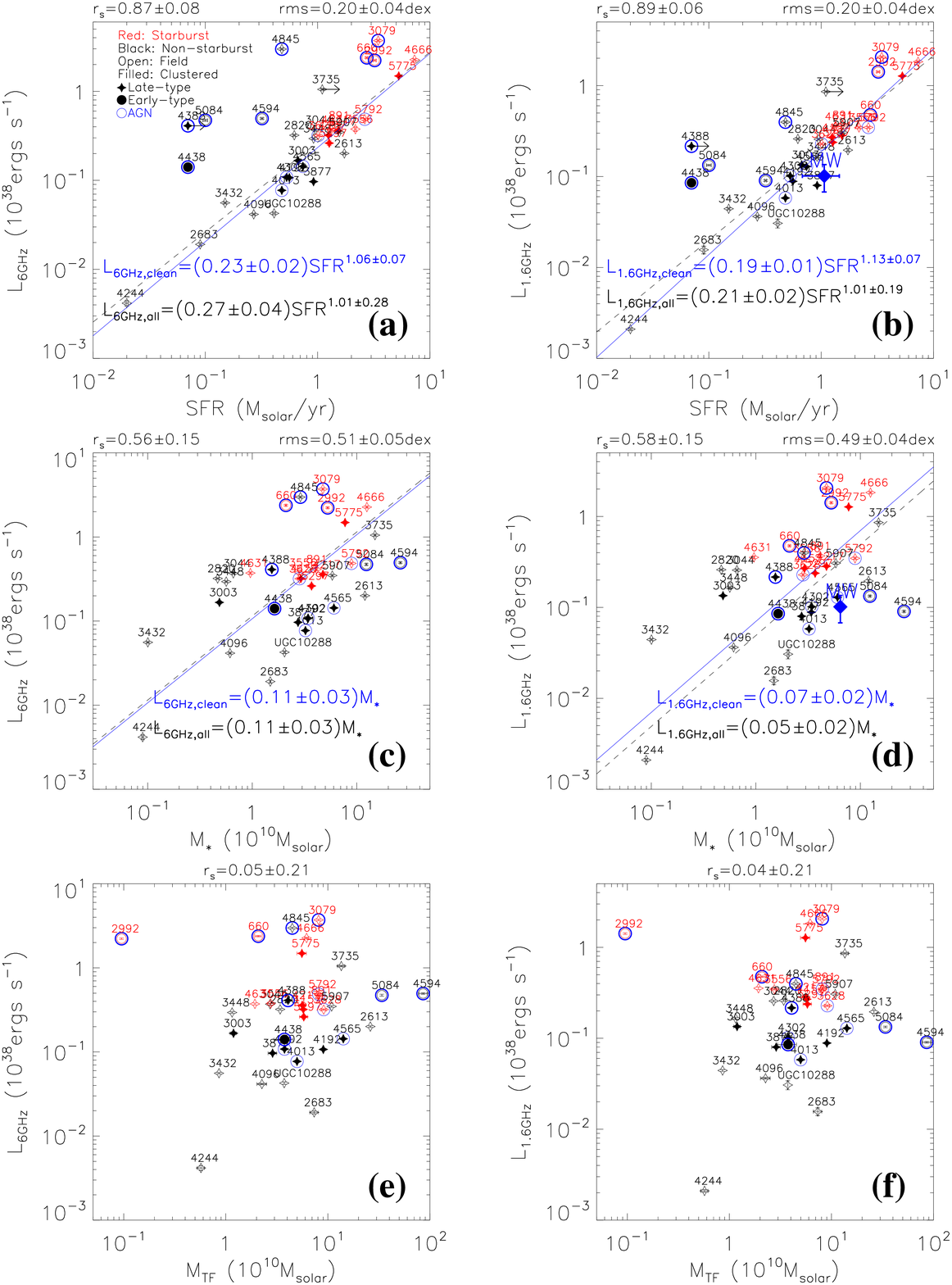,width=0.95\textwidth,angle=0, clip=}
\caption{The C-band (6~GHz) and L-band (1.6~GHz) radio continuum luminosities plotted against the SFR, $M_*$, and $M_{\rm TF}$ of the CHANG-ES galaxies. Different subsamples are plotted with different symbols, as denoted in the upper left corner of panel~(a). The blue diamond in panels (b) and (d) denotes the Milky Way (MW). Names of the galaxies are denoted beside the data points. We show above each panel the Spearman's rank order correlation coefficient ($r_s$) of each relation for galaxies whose radio fluxes are not highly affected by the presence of a radio-bright AGN. We further fit the $L_{\rm radio}-{\rm SFR}$ and $L_{\rm radio}-M_*$ relations for these galaxies alone (the blue solid lines in panels a-d) and present the best-fit relations as well as the $1-\sigma$ errors of the model parameters on the figures. The $rms$ of the data points around the best-fit relations are presented at the top right of each panel. For comparison, we also fit the relations for the whole sample including the highly AGN-contaminated galaxies, present the best-fit relations in black, and plot them as black dashed lines.}
\label{Fig:LradioGalPara}
\end{center}
\end{figure*}

In Fig.~\ref{Fig:LradioGalPara}, we plot the radio continuum luminosities in C- and L-bands against various galaxy properties (SFR, $M_*$, $M_{\rm TF}$). The Milky Way is also included in Fig.~\ref{Fig:LradioGalPara}b,d for comparison, but it is not included in any quantitative calculations, such as the fitting of the relations and the calculation of $r_s$ and $rms$. Galaxies with extremely radio bright AGN (symbols surrounded with a thick blue circle in the figures and listed as ``ERB'' in the last column of Table~\ref{Table:GalPara}) are also excluded from these quantitative calculations. In general, there are tight, weak, and no correlations between the radio luminosities and the SFR, $M_*$, and $M_{\rm TF}$, respectively. The Milky Way is consistent with other galaxies, but may be slightly less luminous in L-band. Most of the ``ERB'' galaxies are significant outliers in Fig.~\ref{Fig:LradioGalPara}a,b, with unusually high radio continuum luminosities.

The $L_{\rm radio}-{\rm SFR}$ correlations shown in Fig.~\ref{Fig:LradioGalPara}a,b are so tight that we can determine the logarithmic slopes of the relations. As the SFR adopted here is linearly related to the IR luminosity ($L_{\rm IR}$; \citealt{Jarrett13}), the slopes of the $L_{\rm radio}-{\rm SFR}$ relations ($L_{\rm radio}$ represents the C- and L-band luminosities $L_{\rm 6GHz}$ and $L_{\rm 1.6GHz}$) are also representative of the slopes of the $L_{\rm radio}-L_{\rm IR}$ relations which have been extensively studied in the literature (e.g., \citealt{Helou85,Yun01,Bell03,Vlahakis07,Basu15a}). The best-fit $L_{\rm radio}-{\rm SFR}$ relations are consistent with a linear slope within 1~$\sigma$ for C-band and a significantly super-linear slope at $\sim2~\sigma$ confidence level for L-band (errors are quoted at 1~$\sigma$ confidence level):
\begin{equation}
\begin{aligned}\label{equi:LCbandSFR}
L_{\rm 6GHz}/{10^{38}\rm ergs~s^{-1}}&=(0.233\pm0.022)\times\\
&({\rm SFR}/{\rm M_\odot~yr^{-1}})^{1.057\pm0.075}
\end{aligned}
\end{equation}
and
\begin{equation}
\begin{aligned}\label{equi:LLbandSFR}
L_{\rm 1.6GHz}/{10^{38}\rm ergs~s^{-1}}&=(0.188\pm0.012)\times\\
&({\rm SFR}/{\rm M_\odot~yr^{-1}})^{1.132\pm0.067}.
\end{aligned}
\end{equation}
The tight correlations between $L_{\rm radio}$ and SFR indicate that young stars should be the major energy source of the radio continuum emission in most of the galaxies, regardless of its thermal or non-thermal origin. In addition, the weak and largely scattered $L_{\rm radio}-M_*$ correlations (also consistent with a linear slope) indicate that old stars traced by near-IR (K-band) emission could also contribute. 

The tight correlation between $v_{\rm rot}$ and $M_*$ (Fig.~\ref{Fig:StarburstDefi}b) indicates that $M_{\rm TF}$ should be a good tracer of the mass of the galaxies. However, there is no correlation between $L_{\rm radio}$ and $M_{\rm TF}$, which apparently indicates that the tight $L_{\rm radio}-{\rm SFR}$ and possibly the weak $L_{\rm radio}-M_*$ correlations are not only a scaling effect in the sense that larger galaxies have more of everything. Although in most of the cases the distribution of individual galaxies are quite similar on the $L_{\rm radio}-M_*$ and $L_{\rm radio}-M_{\rm TF}$ relations, the $L_{\rm radio}-M_{\rm TF}$ relation indeed has much weaker correlation. No matter we include or remove the most significant outlier NGC~2992, which has a clearly warped disk potentially affecting the $v_{\rm rot}$ measurement, the correlation coefficient $r_{\rm s}$ of the $L_{\rm radio}-M_{\rm TF}$ relation is close to 0.

There are several possible reasons which could enlarge the scatter of the $L_{\rm radio}-M_{\rm TF}$ relation. \emph{First}, the inclination correction for $v_{\rm rot}$, which is a key to calculate $M_{\rm TF}$, is based on the optical inclination angle $i$. $i$ is for the stellar component (morphologically determined) and is not necessarily the same as the inclination angle of the gas component (often kinematically determined). Such a difference could add scatter to the derived $M_{\rm TF}$. \emph{Second}, $M_{\rm TF}$ is being derived from a mean relation (the baryonic Tully-Fisher relation with the stellar mass measured from the K-band luminosity; Fig.~\ref{Fig:StarburstDefi}b) with a typical scatter of $\gtrsim0.1\rm~dex$ \citep{Bell01}, which could add some scatter to the $L_{\rm radio}-M_{\rm TF}$ relations. \emph{Third}, $M_{\rm TF}$ includes the contribution from the H{\small~I} gas, which could contribute to the total gravity (in fact, $M_{\rm TF}$ is derived from $v_{\rm rot}$ which is a direct tracer of the total gravity) but is not directly related to the SNe energy injection. The lack of correlation between $L_{\rm radio}$ and $M_{\rm TF}$ thus indicates that the weak $L_{\rm radio}-M_*$ correlations are really produced by processes related to stellar feedback, instead of a purely gravitational effect as is often important in the X-ray emitting hot halos (\citealt{Li13b,Li14}). \emph{Fourth}, the rotation velocity of some galaxies are significantly affected by tidal interaction (e.g., NGC~2992; Fig.~\ref{Fig:StarburstDefi}b). $M_{\rm TF}$ of these galaxies does not trace their mass.

Both young and old stellar populations contribute to the generation of the radio continuum emission. Similar to \citet{Li13b}, we also combine the contributions of young and old stellar populations to the energy budget by defining a total SN energy injection rate of $\dot{E}_{\rm SN (Ia+CC)}=\dot{E}_{\rm SN,Ia}+\dot{E}_{\rm SN,CC}$. In some of the galaxies, especially early-type ones, Type~Ia SNe contribute (apparently) $\gtrsim80\%$ of $\dot{E}_{\rm SN (Ia+CC)}$ (Table~\ref{Table:GalPara}). These galaxies are quiescent in star formation, and Type~Ia SNe should dominate the overall activity of all ISM/CGM phases (e.g., NGC~4594; \citealt{LiZ11}).

The best-fit $L_{\rm radio}-\dot{E}_{\rm SN (Ia+CC)}$ relations are:
\begin{equation}
\begin{aligned}\label{equi:LCbandESN}
L_{\rm 6GHz}/{10^{38}\rm ergs~s^{-1}}&=(0.067\pm0.015)\times\\
&(\dot{E}_{\rm SN (Ia+CC)}/{\rm 10^{41}\rm ergs~s^{-1}})^{1.100\pm0.123}
\end{aligned}
\end{equation}
and
\begin{equation}
\begin{aligned}\label{equi:LLbandESN}
L_{\rm 1.6GHz}/{10^{38}\rm ergs~s^{-1}}&=(0.053\pm0.008)\times\\
&(\dot{E}_{\rm SN (Ia+CC)}/{\rm 10^{41}\rm ergs~s^{-1}})^{1.175\pm0.102},
\end{aligned}
\end{equation}
which also have slightly super-linear slopes at least for L-band ($\sim2~\sigma$ level). We do not find a significant improvement of the correlation by including the contribution from Type~Ia SNe (as compared to the correlation with the SFR). Such an improvement of the correlation of X-ray emission with $\dot{E}_{\rm SN (Ia+CC)}$ is seen by \citet{Li13b}, possibly because there are more early-type galaxies in \citet{Li13b}'s sample.

\begin{figure*}
\begin{center}
\epsfig{figure=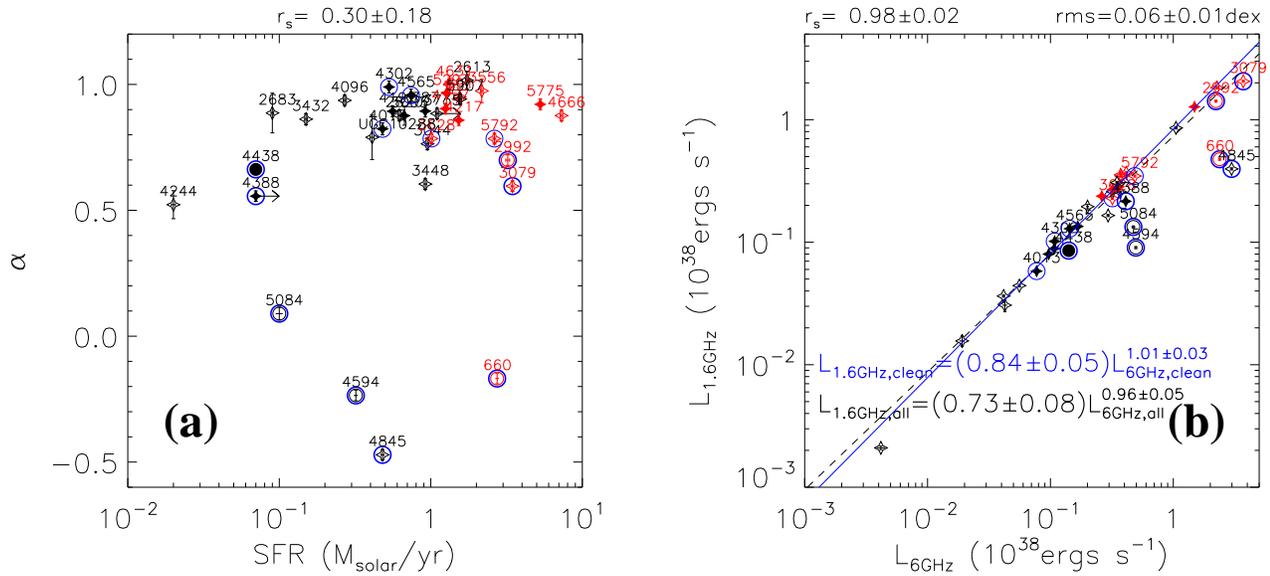,width=1.0\textwidth,angle=0, clip=}
\caption{(a) The C-band (6~GHz) to L-band (1.6~GHz) spectral index ($\alpha$) vs. SFR. (b) Comparing the C-band and L-band luminosities of the galaxies. Symbols are the same as shown in Fig.~\ref{Fig:LradioGalPara}.}\label{Fig:RadioRatioSFR}
\end{center}
\end{figure*}

We further study the integrated radio spectral index between C- and L-bands ($\alpha$; Table~\ref{Table:GalPara}), which is quite indicative of the origin of the radio continuum emission. In general, we do not find a clear trend in any relations between $\alpha$ and other galaxy properties. The positive correlation between $\alpha$ and the SFR is only marginal ($r_s=0.30\pm0.18$; Fig.~\ref{Fig:RadioRatioSFR}a). $\alpha$ typically spans a range of $\sim0.5-1.1$, except for several significant outliers which will be further discussed in \S\ref{subsec:Outliers}. The relatively low scatter of $\alpha$ is also indicated by the extremely tight \emph{linear} correlation between the C-band and L-band luminosities (Fig.~\ref{Fig:RadioRatioSFR}b), which also indicates that the emission in these two bands are produced by closely related processes. 

\begin{figure*}
\begin{center}
\epsfig{figure=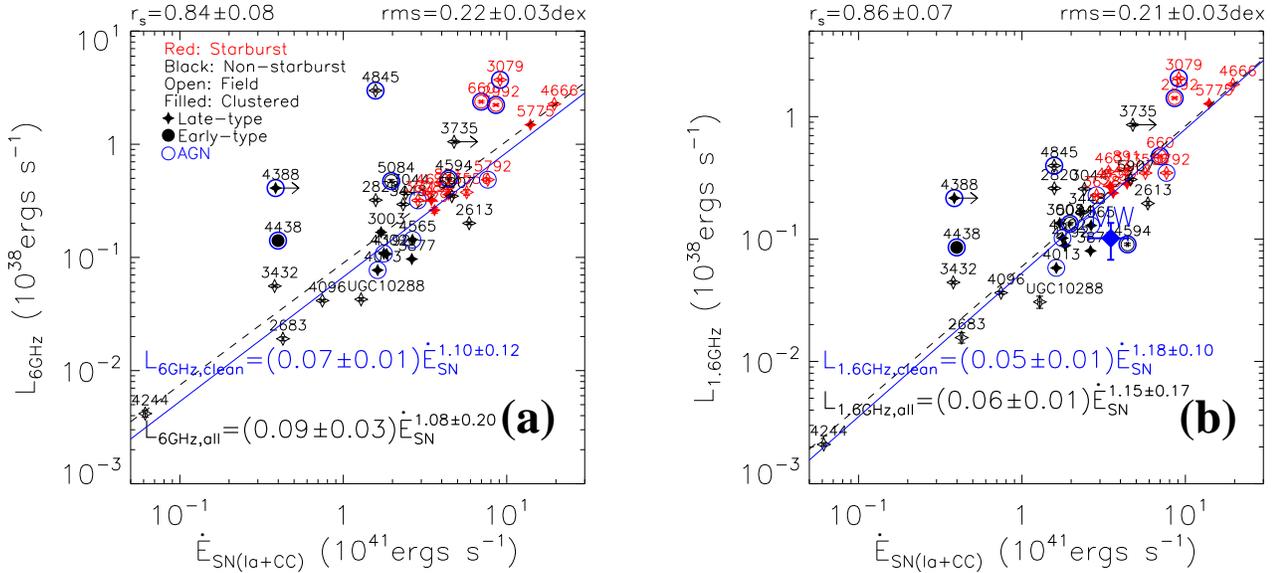,width=1.0\textwidth,angle=0, clip=}
\caption{The C-band (6~GHz) and L-band (1.6~GHz) radio continuum luminosities plotted against the total (core collapsed+Ia) SN energy injection rate ($\dot{E}_{\rm SN (Ia+CC)}$). Symbols are the same as shown in Fig.~\ref{Fig:LradioGalPara}.}\label{Fig:LradioESN}
\end{center}
\end{figure*}

\begin{figure*}
\begin{center}
\epsfig{figure=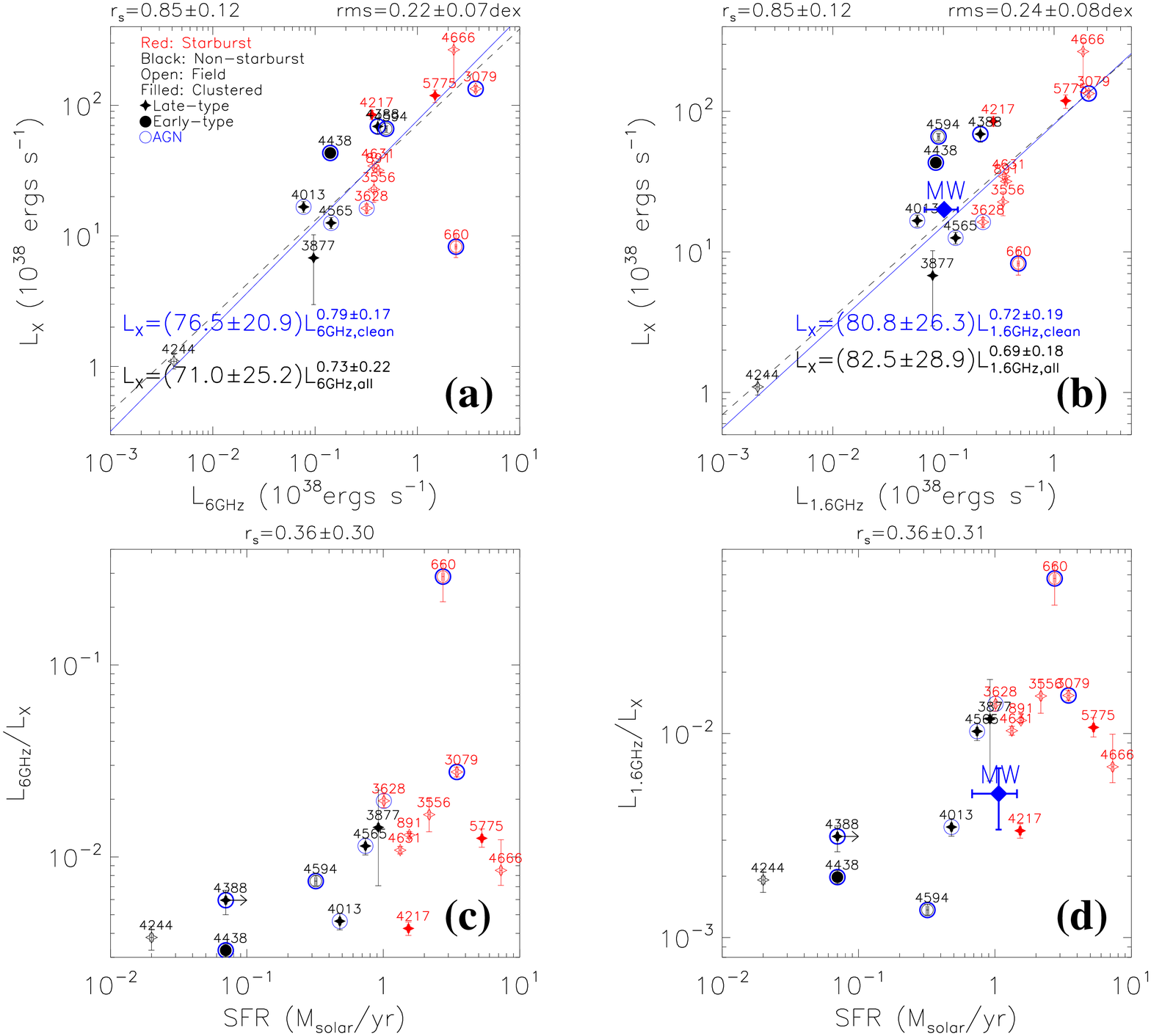,width=1.0\textwidth,angle=0, clip=}
\caption{Comparison of the radio and X-ray emission of the CHANG-ES galaxies. X-ray luminosities are obtained from \citet{Li13a} and scaled to the distances adopted for the CHANG-ES galaxies in the present paper. (a) $L_{\rm X}$ vs. $L_{\rm 6GHz}$; (b) $L_{\rm X}$ vs. $L_{\rm 1.6GHz}$; (c) $L_{\rm 6GHz}/L_{\rm X}$ vs. SFR; (d) $L_{\rm 1.6GHz}/L_{\rm X}$ vs. SFR. Symbols are the same as shown in Fig.~\ref{Fig:LradioGalPara}.}\label{Fig:RadioXray}
\end{center}
\end{figure*}

There is a linear correlation between the point-source-removed truly diffuse halo soft X-ray luminosity and the IR luminosity of \emph{spiral} galaxies (e.g., \citealt{Strickland04,Tullmann06,Li13b}). This correlation is even preserved on sub-galactic scale (e.g., \citealt{Li08}). \citet{Li13b} further find a significantly improved tight linear correlation between the halo X-ray luminosity and the total (CC+Ia) SN energy injection rate of the galaxy. This correlation indicates a roughly constant and extremely low efficiency $\sim(0.4\pm0.1)\%$ of converting the SN energy into the halo X-ray emission. The $L_{\rm radio}-\dot{E}_{\rm SN (Ia+CC)}$ relations (Fig.~\ref{Fig:LradioGalPara}, Eqs.~\ref{equi:LCbandESN},~\ref{equi:LLbandESN}) have even tighter correlation and less scatter but the slopes are likely steeper than the $L_{\rm X}-\dot{E}_{\rm SN (Ia+CC)}$ relation (assumed to be linear in \citealt{Li13b}). The different $L_{\rm radio}-\dot{E}_{\rm SN (Ia+CC)}$ and $L_{\rm X}-\dot{E}_{\rm SN (Ia+CC)}$ relations indicate non-linear $L_{\rm radio}-L_{\rm X}$ relations and higher $L_{\rm radio}/L_{\rm X}$ ratio at higher SFR. These trends are revealed in Fig.~\ref{Fig:RadioXray}, in panels (b) and (d) in which the Milky Way is also consistent with other galaxies.

The best-fit $L_{\rm 6GHz}-L_{\rm X}$ and $L_{\rm 1.6GHz}-L_{\rm X}$ relations are:
\begin{equation}
\begin{aligned}\label{equi:LXLCband}
L_{\rm X}/{10^{38}\rm ergs~s^{-1}}&=(76.5\pm20.9)\times\\
&(L_{\rm 6GHz}/{10^{38}\rm ergs~s^{-1}})^{0.79\pm0.17}
\end{aligned}
\end{equation}
and
\begin{equation}
\begin{aligned}\label{equi:LXLLband}
L_{\rm X}/{10^{38}\rm ergs~s^{-1}}&=(80.8\pm26.3)\times\\
&(L_{\rm 1.6GHz}/{10^{38}\rm ergs~s^{-1}})^{0.72\pm0.19},
\end{aligned}
\end{equation}
with sub-linear slopes at $>1~\sigma$ confidence levels. However, we caution that the best-fit slopes of the $L_{\rm radio}-L_{\rm X}$ relations are highly affected by the faintest galaxy NGC~4244. If we remove this galaxy, the best-fit slopes will be consistent with linear or remain sub-linear for non-AGN galaxies or the whole sample, respectively. The scatter will be larger, however, with the 1~$\sigma$ error of the slopes having typical values of $\sim0.3$. There are also weak correlations between the $L_{\rm radio}/L_{\rm X}$ ratio and the SFR (Fig.~\ref{Fig:RadioXray}c,d). All the relations shown in Fig.~\ref{Fig:RadioXray} indicate that higher-SFR galaxies tend to convert a larger fraction of their SNe energy into non-thermal radio continuum emission than into thermal soft X-ray emission.

\subsection{Significant outliers}\label{subsec:Outliers}

In general, we do not find significant systematic differences between different pre-defined subsamples in the above statistical analysis. However, some galaxies show significantly different properties from others, indicating that their radio emission must have different origins. The nuclei of most of these galaxies are classified as ``ERB'', apparently indicating that it is the AGN or the extended features related to the AGN which produce their unusual properties (e.g., NGC~4845; \citealt{Irwin15}). In this section, we further discuss some additional peculiar properties of these outliers which may affect their multi-wavelength appearance.

\emph{NGC~660} is unusually luminous in C-band (Figs.~\ref{Fig:LradioGalPara}a, \ref{Fig:LradioESN}a) and faint in X-ray (Fig.~\ref{Fig:RadioXray}), while its integrated radio spectral index is also extremely flat and even negative (Fig.~\ref{Fig:RadioRatioSFR}a). The nucleus of this galaxy underwent a GHz-peaked outburst during our \emph{JVLA} D-configuration observations \citep{Argo15}, which will be further discussed in a separate paper. The C-band in-band spectral index map of NGC~660 shows a well-resolved jet-like feature, which also has negative spectral index in the southeast half (\citealt{Wiegert15}; note that the definition of $\alpha$ in this paper corresponds to $-\alpha$ in \citealt{Wiegert15}). Therefore, it is most likely that the AGN and jet is responsible for the high luminosity in C-band and the inverted radio spectral index in the integrated spectrum of this galaxy. NGC~660 is also the galaxy with the lowest X-ray radiation efficiency in \citet{Li13b}'s sample of 53 nearby edge-on galaxies (almost one order of magnitude lower than the best-fit value). Such a low X-ray radiation efficiency is not yet fully understood, because the published X-ray observations of this galaxy by \emph{Chandra} and \emph{XMM-Newton} are rather poor and cannot detect the very extended component in the halo. NGC~660 is an interacting galaxy with a significant fraction of its cold gas widely distributed along the highly warped polar ring \citep{Combes92}. This may help to explain its relative faintness in X-ray, which could be caused by the widely distributed star formation (possibly in the polar ring, which is \emph{not} detected in X-ray; \citealt{Li13a}) and the low hot gas radiative cooling rate in the center because of the lack of cool gas there.

\emph{NGC~4388} and \emph{NGC~4438} are unusually luminous in both C- and L-bands on the $L_{\rm radio}-{\rm SFR}$ and $L_{\rm radio}-\dot{E}_{\rm SN (Ia+CC)}$ relations (Figs.~\ref{Fig:LradioGalPara}a,b, \ref{Fig:LradioESN}), but are quite normal in radio spectral index (Fig.~\ref{Fig:RadioRatioSFR}), and relatively X-ray bright (compared to radio; Fig.~\ref{Fig:RadioXray}. NGC~4438 is also the galaxy with the highest X-ray radiation efficiency in \citealt{Li13b}'s sample). These two galaxies are near the center of the Virgo cluster and show significant tidal interactions with one of the most massive Virgo cluster members, M~86 \citep{Kenney08}. Their diffuse X-ray emission shows clearly distorted structures \citep{Li13a}. Furthermore, NGC~4438 is also found in our \emph{JVLA} observations to have a single-sided radio lobe, and the radio lobe shows strikingly similar fine structure to the H$\alpha$ and X-ray filaments \citep{Machacek04,Kenney08,Li13a}. Therefore, in addition to the presence of the AGN, which is likely the major source of the enhanced radio emission, it is very likely that some environmental effects, such as ram-pressure stripping and/or compression, are affecting both the radio and X-ray properties of these two clustered galaxies (e.g., \citealt{Murphy09,Ehlert13}).

\emph{NGC~4594} and \emph{NGC~5084} are clearly more radio luminous than the best-fit $L_{\rm radio}-{\rm SFR}$ relations (Fig.~\ref{Fig:LradioGalPara}a,b; NGC~4594 is consistent with the best-fit $L_{\rm 1.6GHz}-{\rm SFR}$ relation), but are consistent with other galaxies on the $L_{\rm radio}-\dot{E}_{\rm SN (Ia+CC)}$ relations (Fig.~\ref{Fig:LradioESN}). They both have flat radio spectra (Fig.~\ref{Fig:RadioRatioSFR}), possibly caused by the presence of a low luminosity AGN (e.g., \citealt{Hada13}). NGC~4594 and NGC~5084 are both very massive isolated early-type spiral galaxies. They also have the highest $\dot{E}_{\rm SN,Ia}/\dot{E}_{\rm SN (Ia+CC)}$ ratio among the sample galaxies ($\gtrsim80\%$, Table~\ref{Table:GalPara}). Therefore, it is very likely that a large fraction of the radio continuum emission is contributed by Type~Ia SNe originating from old stellar populations (e.g., \citealt{LiZ11}).

\section{Discussion}\label{sec:Discussions}

Some results from in the above sections, such as the slopes of the $L_{\rm radio}-L_{\rm IR}$ and the $L_{\rm X}-L_{\rm radio}$ relations and the radio spectral index, are of great importance for understanding the origin, propagation, and energy loss of CR particles, as well as how SNe distribute their energy into different phases on galactic scales. 

\subsection{Compare the normalization of the radio-IR relationship to previous results}\label{subsec:DiscussRadioIRNormalization}

For the convenience of comparison with similar studies of the radio-IR relationship in the literature, we convert Eqs.~\ref{equi:LCbandSFR} and \ref{equi:LLbandSFR} to the form of:
\begin{equation}
\begin{aligned}\label{equi:ICbandSFR}
I_{\rm 6GHz}/{\rm W~Hz^{-1}}&=(1.17\pm0.11)\times10^{21}\\
&({\rm SFR}/{\rm M_\odot~yr^{-1}})^{1.057\pm0.075}
\end{aligned}
\end{equation}
and
\begin{equation}
\begin{aligned}\label{equi:ILbandSFR}
I_{\rm 1.6GHz}/{\rm W~Hz^{-1}}&=(3.76\pm0.24)\times10^{21}\\
&({\rm SFR}/{\rm M_\odot~yr^{-1}})^{1.132\pm0.067},
\end{aligned}
\end{equation}
where $I_{\rm 6GHz}$ and $I_{\rm 1.6GHz}$ are the spectral luminosities in units of $\rm W~Hz^{-1}$. 

The normalization of Eq.~\ref{equi:ILbandSFR} is typically 2-3 times higher than those obtained in the literature based on spatially resolved observations of face-on galaxies. For example, \citet{Murphy11} adopted a radio-SFR relation of $I_{\rm 1.4GHz}/{\rm W~Hz^{-1}}=1.57\times10^{21}{\rm SFR}/{\rm M_\odot~yr^{-1}}$ (converted from their Eq.~17), while \citet{Heesen14} obtained a similar relation of $I_{\rm 1.4GHz}/{\rm W~Hz^{-1}}=1.33\times10^{21}{\rm SFR}/{\rm M_\odot~yr^{-1}}$ in their Eq.~1, which is a direct conversion from \citet{Condon02}'s relation by adopting a broken power law initial mass function (IMF) from \citet{Calzetti07}. On the other hand, \citet{Yun01} obtained an integrated (for the whole galaxy) radio-SFR relation of $I_{\rm 1.4GHz}/{\rm W~Hz^{-1}}=1.69\times10^{21}{\rm SFR}/{\rm M_\odot~yr^{-1}}$ in their large sample of 1809 galaxies. Considering the difference in IMF, their normalization should be $2.57\times10^{21}$, between our value and those from spatially resolved observations of face-on galaxies. The difference between $I_{\rm 1.4GHz}$ and the $I_{\rm 1.6GHz}$ in our Equ.~\ref{equi:ILbandSFR} should be negligible. Therefore, compared to the papers on individual star formation regions in face-on galaxies, our integrated relation of edge-on galaxies has either over-predicted the radio flux or under-predicted the SFR.

We carefully checked our data calibration processes. In Paper~IV and the present paper, the SFR is estimated based on the \emph{WISE} 22$\rm~\mu m$ luminosity using \citet{Jarrett13}'s relation, which adopts the IMF described in \citet{Rieke09} (\S\ref{subsec:GalPara}). There are 16 galaxies in the CHANG-ES sample overlapping with \citet{Li13a}'s sample, where they estimate the SFR from the \emph{IRAS} total-IR luminosity using \citet{Kennicutt98}'s relation. The total-IR based SFR in \citet{Li13a} is on average $\sim2.3$ times of the mid-IR based SFR in this paper, after accounting for the different IMF adopted in these two papers, using the same definition of total-IR luminosity as \citet{Irwin12a}, and adopting the more recent \emph{IRAS} fluxes (\citealt{Sanders03}, while \citealt{Li13a} adopted the \emph{IRAS} fluxes from \citealt{Fullmer89}, which are compiled before the final calibration of the \emph{IRAS} Level~1 Archive in 1990 May). Our radio flux is in general consistent with those collected from the literature in \citet{Irwin12a}, except for several galaxies hosting an AGN in the outburst stage during the observations (e.g., NGC~4845, \citealt{Irwin15}). Therefore, the discrepancy in the radio-IR relationships between our Eq.~\ref{equi:ILbandSFR} and some papers could be attributed to several processes:

\emph{Firstly, strong absorption by dust grains in the edge-on case may even affect the IR emission.} The discrepancy between our radio-IR relation and that from \citet{Yun01} (which is also integrated for the whole galaxy) cannot be a result of the diffusion of CR electrons (with a typical length scale of $\sim\rm kpc$ at L-band, e.g., \citealt{Berkhuijsen13}). As already pointed out in \S\ref{subsec:GalPara}, the average total-IR to mid-IR luminosity ratio of our sample is $\sim$18\% to 60\% higher than those in \citet{Rieke09} or \citet{Calzetti10}, which may be caused by the strong dust absorption in our edge-on galaxies. Although in many cases, the total-IR and mid-IR emission can be regarded as extinction free, this may not be true at high inclination angle of the galactic disk. The $\sim2.3$ times difference between the mid-IR and total-IR based SFR may explain the discrepancy in the radio-IR relation between our result and \citet{Yun01}'s. However, we caution that \citet{Yun01} adopted a far-IR based SFR which could be biased by the relatively strong contribution from cold dust heated by low-mass stars (see below). Furthermore, such a discrepancy in mid-IR and total-IR based SFRs also exists in other papers, e.g., $\sim1.8$ times for \citet{Rieke09} and $\sim1.4$ times for \citet{Calzetti10}. A better estimate of the SFR should be based on an SED fit from radio to IR, especially by including the extinction free and thermal dominated 33~GHz emission (e.g., \citealt{Murphy11}).

\emph{Secondly, heating of dust by non-ionizing photons from low mass stars could contribute to the IR emission.} Non-ionizing photons from low mass stars could also contribute to the heating of the circumstellar dust so the IR emission (e.g., \citealt{Temi09}). This IR emission is not related to the current star formation, so could bias the estimation of SFR based on any sort of dust reprocessed IR emission. \citet{Murphy11} found that the IR-based SFR in the nuclear region of NGC~6946 is clearly over-estimated, probably because there are more old stars there. When estimating the SFR based on the \emph{WISE} 22$\rm~\mu m$ luminosity, we have removed IR-bright AGN and only include the IR emission from the disk. This process may also reduce the contamination from old stars in the galactic bulge, thus results in a lower SFR.

\subsection{The calorimeter condition and its effect on the radio-IR slope}\label{subsec:DiscussCalorimeterCondition}

We next roughly estimate the calorimeter conditions for CR leptons in galactic disks and halos. The synchrotron cooling timescale for CR electrons is $\tau_{\rm syn}\sim10^9{\rm~yr}~B^{-1.5}\nu^{-0.5}$, where $B$ is the magnetic field strength measured in $\rm\mu G$, and $\nu$ is the frequency of synchrotron emission measured in GHz \citep{Beck13}. For a rough estimation, we assume the average magnetic field of a spiral galaxy is $B\sim10\rm~\mu G$. For C- and L-bands, the resultant $\tau_{\rm syn}$ are then $\sim1.3\times10^7\rm~yr$ and $\sim2.5\times10^7\rm~yr$, respectively. For comparison, if we assume the CR electrons are entrained in a galactic wind with a typical velocity of $\sim300\rm~km~s^{-1}$, the typical escape timescale will be $\sim10^6\rm~yr$ and $3\times10^7\rm~yr$ for a galactic disk with $\sim300\rm~pc$ scale height and a galactic halo with a typical size of $\sim10\rm~kpc$, respectively. Therefore, a galactic disk is typically \emph{not} a calorimeter for CR electrons entrained in galactic outflows. On the other hand, a galactic halo is closer to a CR electron calorimeter, with the real condition depending on the specific parameters. These estimates highlight the importance of including the galactic halo (with a typical extension of $\lesssim10\rm~kpc$ for the CHANG-ES galaxies based on the D-configuration observations; \citealt{Wiegert15}) into the energy budget of CRs, which could only be done with deep radio continuum observations such as those being carried out in the CHANG-ES program.

\citet{Volk89} developed a calorimeter model in order to explain the tight radio-IR correlation. This model assumes that each galaxy act as a ``calorimeter'' for both the relativistic electrons and the UV photons which heat the surrounding dust and are re-radiated in IR. The relativistic electrons lose energy mainly via synchrotron and inverse Compton (IC) processes. Therefore, in \citet{Volk89}'s model, the only factor that could modify the slope of the $L_{\rm radio}-L_{\rm IR}$ relation or the proportionality of the non-thermal radio emissivity to the SN rate is the average energy density ratio of the radiation and magnetic field $\bar{U}_{\rm rad}/\bar{U}_{\rm B}$ ($L_{\rm radio}/\dot{E}_{\rm SN}\propto(1+\bar{U}_{\rm rad}/\bar{U}_{\rm B})^{-1}$). As the total (or random) magnetic field strength increases much slower than $\bar{U}_{\rm rad}$ with the SFR, and the ordered magnetic field even shows no clear increase (e.g., \citealt{Stil09,Krause11}), the $\bar{U}_{\rm rad}/\bar{U}_{\rm B}$ ratio is expected to be higher for high-SFR galaxies (e.g., \citealt{Thompson06}). Therefore, the calorimeter model predicts sub-linear $L_{\rm radio}-{\rm SFR}$ or $L_{\rm radio}-\dot{E}_{\rm SN (Ia+CC)}$ relations, or less energy of CR electrons dissipated via synchrotron emission at higher SFR. The calorimeter model could successfully reproduce the global multi-wavelength CR-related emission in the extended (e.g., $\sim10\rm~kpc$) Milky Way halo \citep{Strong10}.

A non-calorimeter model discussed in \citet{Niklas97a} predicts a logarithm radio-IR slope of $\sim1.3$. This model assumes equipartition between the energy densities of the CR and magnetic field, and adopts the \emph{observed} dependencies of the SFR and the strength of the equipartition magnetic field on the density of the cool gas, as well as the slope of the radio synchrotron spectrum. In non-calorimeter models such as the \citet{Niklas97a} model, it is natural to have non-calorimeter conditions on small scales or at low and moderate SFRs. In this case, most of the CR electrons escape from the galactic disk which accounts for a dominant fraction of the detected non-thermal radio emission. The production rate of CR electrons has a stonger dependence on the SFR ($\propto{\rm SFR}$) than the magnetic field energy density ($\propto B^2 \propto {\rm SFR}^{0.69}$; \citealt{Niklas97a}). Therefore, although the synchrotron cooling timescale decreases with increasing SFR (e.g., \citealt{Lacki10}), it is still expected that the escape probability of CR electrons out of the galactic disk increase with the SFR (however, see an opposite trend predicted by \citealt{Chi90}). However, unlike calorimeter models in which the non-thermal radio emission does not depend significantly on the magnetic field strength, the non-calorimeter energy equipartition model of \citet{Niklas97a} predicts a strong dependence of the non-thermal radio emission on the magnetic field strength ($L_{\rm radio}\propto B^4 \propto {\rm SFR}^{1.37}$), which compensates the weak positive dependence of the escape probability on the SFR. In addition, it is not clear if the velocity of galactic wind depends significantly on the SFR or not. If so, the escaping timescale should be shorter in high SFR galaxies, and the dependence of the escape probability on the SFR should be weaker. A combination of these effects naturally produces a super-linear radio-IR slope.

In conclusion, the radio-IR slope depends on the SFR. \emph{At low and moderate SFR}, the magnetic field is low and the synchrotron radiative cooling timescale is long, so most of the CR electrons could escape from the galactic disk. As predicted by \citet{Niklas97a}'s non-calorimeter model, there is a super-linear radio-IR relation in this escape-dominated regime. On the other hand, \emph{at high SFR}, the magnetic field (and also the radiation field) is high enough that most of the CR electrons lose their energy via synchrotron radiation or other cooling processes (e.g., IC) close to the disk. In this loss-dominated regime, a calorimeter condition is satisfied and the radio-IR relation is almost linear, or if the IC cooling is dominant, the relation can be even sub-linear. We do \emph{not} find significant evidence for the flattening of the radio-IR relation at high SFR within our sample. This is either because the SFRs of our sample galaxies are not as high as required by the loss-dominated regime, or because the escape timescale also decreases with the SFR as a result of the possibly increasing wind velocity with the SFR (e.g., \citealt{Krause09}).

\subsection{The observed slopes of the $L_{\rm radio}-{\rm SFR}$ and $L_{\rm radio}-\dot{E}_{\rm SN (Ia+CC)}$ relations}\label{subsec:DiscussSuperLinearSlope}

\begin{figure}
\begin{center}
\epsfig{figure=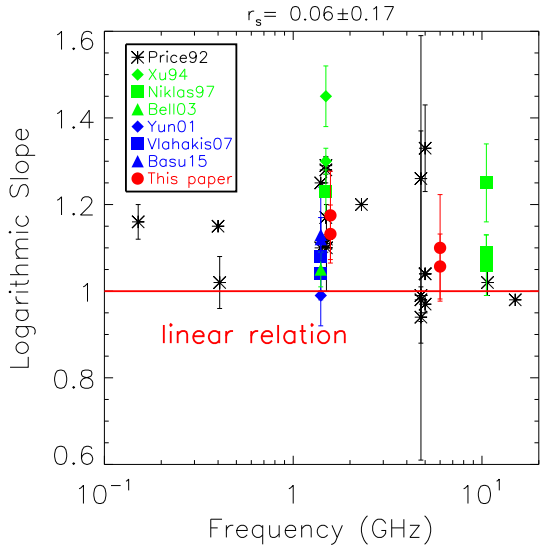,width=1.0\linewidth,angle=0, clip=}
\caption{The logarithmic slopes of the $L_{\rm radio}-L_{\rm IR}$ relations obtained from the present paper (red filled circles; for both the $L_{\rm radio}-{\rm SFR}$ and $L_{\rm radio}-\dot{E}_{\rm SN (Ia+CC)}$ relations of Eqs.~\ref{equi:LCbandSFR}, \ref{equi:LLbandSFR}, \ref{equi:LCbandESN}, \ref{equi:LLbandESN}) and literature plotted against the center frequency of the radio observations. The red solid line marks the slope of a linear relation.}\label{Fig:RadioIRSlopes}
\end{center}
References of the data are denoted in the inset panel: Price92: early measurements collected by \citet{Price92}, including their own measurements; Xu94: \citet{Xu94}; Niklas97: \citet{Niklas97b}; Yun01: \citet{Yun01}; Bell03: \citet{Bell03}; Vlahakis07: \citet{Vlahakis07}; Basu15: \citet{Basu15a}.
\end{figure}

In Fig.~\ref{Fig:RadioIRSlopes}, we compare the logarithmic slopes of the $L_{\rm radio}-{\rm SFR}$ and $L_{\rm radio}-\dot{E}_{\rm SN (Ia+CC)}$ relations (Eqs.~\ref{equi:LCbandSFR}, \ref{equi:LLbandSFR}, \ref{equi:LCbandESN}, \ref{equi:LLbandESN}) to some archival measurements of the slopes of the $L_{\rm radio}-L_{\rm IR}$ relations. We caution that these archival measurements are obtained with different samples, different radio observing depth and array configurations (or single-dish observations), different bands for IR luminosities, and sometimes even different redshifts. Therefore, the figure cannot be used for quantitative calculations.

Our measurements based on the CHANG-ES data show a clearly steeper slope in L-band than in C-band, and a significantly super-linear $L_{\rm radio}-L_{\rm IR}$ relation in L-band. However, the super-linearity in C-band is quite marginal. The large scatter of the measurements obtained from different papers also prevents us from finding a clear trend of steeper $L_{\rm radio}-L_{\rm IR}$ relations (larger slope) at lower frequencies as claimed by \citet{Price92}. Such a trend is expected if there is a lower fraction of non-thermal radio emission at high frequencies and/or a faster synchrotron energy loss and more calorimeter-like condition for higher energy CR electrons.

The discussions in \S\ref{subsec:DiscussCalorimeterCondition} assume the radio continuum emission is mainly produced by the synchrotron process. However, the radio continuum emission at C- and L-bands actually include both thermal and non-thermal components (e.g., \citealt{Price92,Dumke00,Chyzy07,Krause09,Irwin12b}). The thermal radio emission is expected to be linearly dependent on the thermal IR emission produced by the dust-reprocessed high energy photons from young stars (e.g., \citealt{Price92,Xu94,Niklas97b}). As a result, in addition to a more calorimeter-like condition in C-band, the flatter radio-IR relations in C-band than in L-band (Fig.~\ref{Fig:RadioIRSlopes}) could also be partially caused by the larger contribution of thermal emission in C-band.

Another possibility to explain the high $L_{\rm radio}/{\rm SFR}$ or $L_{\rm radio}/\dot{E}_{\rm SN (Ia+CC)}$ ratios of high-SFR galaxies are the additional CR lepton sources in these galaxies. These additional CR leptons may come from the secondary electrons/positrons produced by the decay of charged pions ($\pi^+$ or $\pi^-$), which are produced in the interaction between CR hadrons and the ISM (e.g., \citealt{Strong07,Lacki10}). In a galaxy similar to the Milky Way, typically $\sim (10-20)\%$ of the synchrotron energy loss in the radio band is caused by the secondary electrons/positrons \citep{Strong10}. However, in intense star forming galaxies, the ISM density is much higher, and the galaxy becomes a better calorimeter for CR protons due to frequent proton (CR)-proton (ISM) interaction and strong pion decay. In these galaxies, the secondary electrons could even dominate the primary ones (e.g., \citealt{DomingoSantamaria05,Lacki10}). Therefore, intense star forming galaxies possibly have a higher non-thermal radio flux density per unit SFR or $\dot{E}_{\rm SN (Ia+CC)}$ than inactively star-forming ones. 

\subsection{Radio spectral indices}\label{subsec:DiscussRadioSpecShape}

The spatially-integrated radio spectral index of the whole galaxy as presented in Fig.~\ref{Fig:RadioRatioSFR}a could be influenced by various processes. In the absence of a radio-bright AGN, there are three major components of the galactic radio continuum emission \citep{Lisenfeld00}: the thermal emission from H{\small~II} regions, typically having a flat spectrum ($\alpha\sim0.1$, e.g., \citealt{Price92}); the synchrotron emission from discrete SNRs with an almost constant spectral index of $\alpha\sim0.5$ \citep{Green14}; and the diffuse synchrotron emission from the CR electrons spread over the entire galaxy, where the spectral index varies in a large range, typically from 0.5 to 1.1 \citep{Lisenfeld00}. 

The integrated radio spectral index is quite indicative of the origin of the radio continuum emission. The steep spectrum or the large $\alpha$ ($\sim0.5-1.1$) of most of the sample galaxies is very consistent with a non-thermal origin of the radio continuum emission from the diffuse synchrotron component. However, the galactic scale diffuse synchrotron emission is affected by many factors (e.g., \citealt{Strong11,Orlando13,Basu15b}), such as the CR lepton spectral index integrated along the line of sight, the type of propagation (diffusion, convection, and re-acceleration processes), the relative importance of CR electron energy loss mechanisms (via synchrotron, inverse Compton, adiabatic loss, bremsstrahlung, and/or ionization), and the confinement of the CR electrons. The resultant spectral index of this component is thus highly variable, and its nature cannot be simply determined based on the spatially integrated spectral index alone. 

On the other hand, a substantial, but not dominant contribution of thermal emission to the radio continuum is \emph{not inconsistent} with the large measured $\alpha$ values. It is found that the relative contributions of the thermal and non-thermal components are always variable among galaxies (e.g., \citealt{Chyzy07,Krause09}), and a significant contribution from the thermal component has been revealed in several galaxies of our sample (e.g., NGC~5907, \citealt{Dumke00}). Furthermore, the in-band spectral index maps of many CHANG-ES galaxies indicate clear spatial variations of $\alpha$ \citep{Wiegert15}, with the radio spectrum often flattening toward the galactic disk, indicating a larger contribution from the thermal emission in the disk or a flattening of the synchrotron spectral index in star-forming regions (e.g., \citealt{Tabatabaei07,Basu13}).

Fig.~\ref{Fig:RadioRatioSFR}a shows only a very weak correlation between $\alpha$ and the SFR ($r_s=0.30\pm0.18$ for all but ``ERB'' galaxies). However, if we remove all the galaxies with AGN (both ``RB'' and ``ERB''), the correlation becomes better ($r_s=0.46\pm0.18$). The scatter in Fig.~\ref{Fig:RadioRatioSFR}a is so large that we cannot accurately determine the $\alpha-{\rm SFR}$ slope, but the weak correlation between $\alpha$ and the SFR is consistent with the difference between the slopes of the $L_{\rm 6GHz}-{\rm SFR}$ and $L_{\rm 1.6GHz}-{\rm SFR}$ relations. Even if we remove all the galaxies with AGN (``ERB''+``RB''), starburst galaxies on average still do not show significantly higher $\alpha$ than non-starburst ones, except for extremely quiescent galaxies such as NGC~4244. Therefore, although we cannot exclude a possible correlation between $\alpha$ and the SFR just based on the current data, it is clear that such a relation cannot be monotonic. There must be some mechanisms limiting the steepening of the radio spectrum at very high SFR.
 
Assuming the radio emission is dominated by the non-thermal component, it is expected that the calorimeter condition could strongly affect the radio spectral index (e.g., \citealt{Lacki10}). For a very simple estimate under the assumption of continuous CR electron injection, by assuming an energy spectrum of CR electrons with a power law index of $p\sim2-2.5$, the radio synchrotron spectrum will have a spectral index of $\alpha\sim1.0-1.2$ under calorimeter conditions. A flatter spectrum may indicate non-calorimeter conditions (e.g., \citealt{Lacki10}). In high-SFR galaxies, the CR electron is often dominated by a single-injection population (e.g., \citealt{Heesen15}). These starburst galaxies have strong magnetic fields, the rapid energy loss of CR electrons therefore produces a steep radio spectrum. This trend is in general consistent with a positive correlation between $\alpha$ and the SFR. However, on the other hand, extreme starburst galaxies often produce strong galactic winds or fountains which could enhance the escape of CR leptons via convection, therefore reducing the radiative loss and flattening the synchrotron spectrum. Furthermore, additional cooling of CR electrons in starburst galaxies, e.g., via ionization, could also flatten their radio spectra at low frequency (e.g., \citealt{Thompson06,Basu15b}). This suggests that extremely active or inactive star forming galaxies tend to have flatter radio spectra, consistent with the trend shown in Fig.~\ref{Fig:RadioRatioSFR}a. 

\subsection{Regulation of SNe energy feedback in radio and X-ray bands}\label{subsec:DiscussRadioXray}

We first estimate the radiation efficiency of CR electrons in radio bands typically dominated by the non-thermal component (e.g., from 10~MHz to 10~GHz). As shown in Eqs.~\ref{equi:LCbandESN} and \ref{equi:LLbandESN}, the SNe radiation efficiency in C- and L-bands (the fraction of SNe energy radiated in C- and L-bands) are $\eta_{\rm C}\sim8\times10^{-5}$ and $\eta_{\rm L}\sim5\times10^{-5}$, respectively. A direct power law interpolation of $\eta_{\rm C}$ and $\eta_{\rm L}$ to the whole radio band dominated by non-thermal emission results in an radio continuum radiation efficiency of $\eta_{\rm radio}\equiv L_{\rm10~MHz-10~GHz}/\dot{E}_{\rm SN (Ia+CC)}\sim8\times10^{-4}$. If we assume a SN could typically convert $\lesssim10\%$ of its mechanical energy into CRs (e.g., \citealt{Kosenko14}), the estimated $\eta_{\rm radio}$ indicates $\sim1\%$ of the CR energy could be converted to radio continuum emission around the whole galaxy (including the halo). Since most of the radio continuum emission is produced by CR electrons, if we assume a calorimeter condition for the observed radio halo, this fraction of $\sim1\%$ is also the fraction of CR energy contained in electrons, and is consistent with the estimates for the Milky Way (e.g., \citealt{Strong10}). In conclusion, SNe could in general provide enough CR electrons responsible for the observed radio continuum emissions from the whole galaxy.

The X-ray properties of galaxies have much larger scatter than the radio properties. This is mainly because a galaxy is typically \emph{not} a ``calorimeter'' for X-ray emitting hot gas at least in the inner halo where most X-ray observations are sensitive enough (e.g., \citealt{Li13a}). On the other hand, a galactic halo is often a fairly good ``calorimeter'' for radio continuum emitting CR electrons as discussed in \S\ref{subsec:DiscussCalorimeterCondition}. This difference results in a large scatter on the X-ray radiation efficiency $\eta_{\rm X}$ ($\equiv L_{\rm X}/\dot{E}_{\rm SN (Ia+CC)}$) and poorly constrained slopes of the X-ray related scaling relations (e.g., Fig.~\ref{Fig:RadioXray}; \citealt{Li13b,Wang15}), but a small scatter in $\eta_{\rm radio}$ and relatively tight $L_{\rm radio}-{\rm SFR}$ and $L_{\rm radio}-\dot{E}_{\rm SN (Ia+CC)}$ relations (Figs.~\ref{Fig:LradioGalPara}a,b, \ref{Fig:LradioESN}). Therefore, our conclusions on the slopes of the scaling relations are mostly based on these tighter relations.

There are two possibilities to explain the trend of the dependence of the radio/soft X-ray luminosities on the SFR or $\dot{E}_{\rm SN (Ia+CC)}$. \emph{Firstly}, as discussed in \S\ref{subsec:DiscussCalorimeterCondition} and \ref{subsec:DiscussSuperLinearSlope}, because of the strong dependence of the radio continuum emission on the magnetic field strength in non-calorimeter models and/or the larger contribution from secondary electrons, there is a higher radio radiation efficiency $\eta_{\rm radio}$ in higher-SFR galaxies. \emph{Alternatively}, the determination of the $L_{\rm X}-\dot{E}_{\rm SN (Ia+CC)}$ slope is less accurate (than the $L_{\rm radio}-\dot{E}_{\rm SN (Ia+CC)}$ or $L_{\rm X}-L_{\rm radio}$ relations), for example, $rms\sim0.5\rm~dex$ over $\sim3\rm~dex$ of $L_{\rm X}$ or $\dot{E}_{\rm SN (Ia+CC)}$ \citep{Li13b}. Therefore, it is also possible that the X-ray radiation efficiency $\eta_{\rm X}$ slightly decreases with increasing $\dot{E}_{\rm SN (Ia+CC)}$. In fact, $\eta_{\rm X}$ spans a large range of $\sim2\rm~dex$, and depends on some other galaxy parameters, indicating that it could be affected by both cool-hot gas interaction and gravitational confinement \citep{Li13b,Wang15}. 

In the inner region of a galactic halo, the diffuse soft X-ray emission is often produced by a galactic superwind, fountain or subsonic outflow (e.g., \citealt{Strickland04,Li13b,Li14}). The galactic superwind could be launched by many mechanisms, such as the energy and momentum deposition by SNe, the radiation pressure, and/or the CRs (e.g., \citealt{Everett08,Hopkins12}). There are two key but poorly constrained parameters which could determine how strong the X-ray emission related to this superwind is: the thermalization efficiency of the SNe energy and the mass loading efficiency into the wind (e.g., \citealt{Chevalier85,Strickland00,Strickland09,Wang15}). \citet{Zhang14} show that if the thermalization efficiency and the mass loading efficiency are both constant, the X-ray luminosity of a galactic wind should be $\propto\rm SFR^2$, much steeper than the observed linear relations (e.g., \citealt{Li13b}). Instead, either or both efficiencies must vary with the SFR. \citet{Zhang14} further show that the mass loading efficiency could decrease significantly in high-SFR galaxies. This result is consistent with the observed anti-correlation between the X-ray radiation efficiency $\eta_{\rm X}$ and the CC SN rate surface density \citep{Li13b,Wang15}, as well as the trend between the X-ray brightness of a sample of earlier-type galaxies and the cool gas content \citep{Li11}. In conclusion, the X-ray radiation efficiency could decrease with increasing SFR, but this is most likely an effect of the decreasing mass loading efficiency, and we do \emph{not} have direct evidence that a galaxy distributes a smaller fraction of its SNe feedback energy into galactic superwinds.

\section{Summary and conclusions}\label{sec:Summary}

It is widely believed that SNe and SNRs are the major sources of CR leptons, which lose their energy via synchrotron radiation and other processes, producing the radio continuum emission in galaxies. The extended radio continuum halo detected in some nearby galaxies strongly suggests that galaxy disks are \emph{not} calorimeters for CR electrons. However, whether or not a galactic \emph{halo} satisfies the calorimeter condition is still under debate and could be affected by many factors. 

Thanks to the high sensitivity and angular resolution of the \emph{JVLA}, we obtain reliable spatially resolved measurements of the \emph{diffuse} radio continuum properties of a flux-limited nearby edge-on spiral galaxy sample (CHANG-ES; \citealt{Irwin12a,Wiegert15}). The CHANG-ES data enable us, for the first time, to include the galactic radio halo into the energy budget for a statistically meaningful sample. In this paper, we perform statistical analyses, comparing the integrated diffuse radio continuum luminosity of the whole galaxy (including both disk and halo) to other galaxy properties. The Milky Way is also included for comparison. Our main results and their scientific implications are summarized below.

(1) We compare the C- and L-band radio continuum luminosities to the SFR (estimated from the spatially-resolved \emph{WISE} mid-IR photometry), stellar mass ($M_*$, estimated from the \emph{2MASS} near-IR photometry), and total baryonic mass ($M_{\rm TF}$, estimated from the gas rotation velocity) of the galaxies. We find tight $L_{\rm radio}-{\rm SFR}$ and weak $L_{\rm radio}-M_*$ correlations, indicating that young stars are the major energy sources of, while old stars could also contribute to, the radio continuum emission. The lack of $L_{\rm radio}-M_{\rm TF}$ correlation further indicates that the $L_{\rm radio}-{\rm SFR}$ and $L_{\rm radio}-M_*$ correlations are not only a scaling effect in the sense that larger galaxies have more of everything. The Milky Way is in general consistent with the best-fit relations of other galaxies within the scatter.

(2) We combine the contributions from the energy injection of CC and Type~Ia SNe ($\dot{E}_{\rm SN (Ia+CC)}$), which are proportional to the SFR and $M_*$, respectively. The tight and low-scatter $L_{\rm radio}-\dot{E}_{\rm SN (Ia+CC)}$ correlations indicate that the radio continuum emission on average accounts for $\sim8\times10^{-4}$ of the total SNe energy injection rate. This radio radiation efficiency is consistent with the expected energy contained in CR electrons, which suggests that the observed radio halo of the CHANG-ES galaxies are very close to a calorimeter of the CR electrons.

(3) The scatter of the $L_{\rm radio}-{\rm SFR}$ and $L_{\rm radio}-\dot{E}_{\rm SN (Ia+CC)}$ relations are so small ($rms\lesssim10\%$ of the logarithmic covering range of the parameters) that we could accurately constrain their logarithmic slopes, which are clearly (at $\sim2~\sigma$ confidence level) super-linear in L-band ($1.132\pm0.067$ and $1.175\pm0.102$ for the $L_{\rm radio}-{\rm SFR}$ and $L_{\rm radio}-\dot{E}_{\rm SN (Ia+CC)}$ relations, respectively) while marginally (within $\sim1~\sigma$) consistent with linear in C-band ($1.057\pm0.075$ and $1.100\pm0.123$ for the $L_{\rm radio}-{\rm SFR}$ and $L_{\rm radio}-\dot{E}_{\rm SN (Ia+CC)}$ relations, respectively). 

(4) The normalization of our $I_{\rm 1.6GHz}/{\rm W~Hz^{-1}}-{\rm SFR}$ relation is typically 2-3 times higher than those obtained in some papers based on spatially resolved observations of face-on galaxies, indicating that we have either over-predicted the radio flux or under-predicted the SFR compared to these works. This discrepancy may plausibly be explained by the enhanced extinction in our edge-on galaxies even in IR, while it could also be affected by the heating of the dust by non-ionizing photons from low-mass stars in the galactic bulge.

(5) The integrated radio spectral index between C- and L-bands is large ($\alpha\sim0.5-1.1$) for most of the sample galaxies, apparently consistent with a non-thermal origin from the galaxy-wide diffuse synchrotron emission in both bands. However, we cannot rule out a considerable contribution from the thermal emission, which may add scatter to the measured $\alpha$ and also contribute to the spatial variation of the in-band spectral index maps as presented in \citet{Wiegert15}. 

(6) Based on the halo diffuse soft X-ray luminosity $L_X$ measured with the \emph{Chandra}, we find sub-linear $L_{\rm X}-L_{\rm radio}$ slopes in both C- and L-bands. Furthermore, the $L_{\rm radio}/L_{\rm X}$ ratio also weakly depends on the SFR. Therefore, there is a faster increase of radio continuum emission than X-ray emission with increasing SFR.

(7) There are several galaxies with radio-bright AGN that they show clearly different radio and/or X-ray properties from other galaxies. They are typically more radio-luminous and have flat radio spectra. In addition to the contamination from the AGN, some of these outliers also exhibit peculiar properties, such as tidal interactions, environmental effects related to the ICM, and large contributions from the feedback of Type~Ia SNe. These unusual properties may further affect their radio and X-ray properties.

Although the overall energy budget indicates that the galactic radio continuum halo detected by the CHANG-ES observations is close to a calorimeter for CR electrons in order of magnitude, the super-linear slopes of the $L_{\rm radio}-{\rm SFR}$ and $L_{\rm radio}-\dot{E}_{\rm SN (Ia+CC)}$ relations (at least in L-band) are inconsistent with the calorimeter model for the disk. Instead, the super-linearity could be naturally reproduced with equipartition non-calorimeter models. These non-calorimeter models predict a strong dependence of the non-thermal radio emission on the magnetic field strength, which depends on the SFR. In addition to a higher radio radiation efficiency, the X-ray radiation efficiency may also slightly decrease in intense star forming galaxies (\citealt{Li13b,Wang15}), helping to produce the sub-linear $L_{\rm X}-L_{\rm radio}$ slopes.

The super-linear radio-IR relation indicates that high-SFR galaxies are closer to calorimeter conditions. This scenario also predicts that the galaxies are better calorimeters for higher energy electrons because their energy loss via synchrotron radiation is faster than for lower energy electrons. Therefore, there should be a flatter radio-IR relation at higher frequencies. Such a trend is revealed by our results in C- and L-bands, but we could not further confirm it by comparing with archival observations, mainly because of the large diversity in various archival observations. Furthermore, different radio-IR slopes in C- and L-bands also predict a weak dependence of the integrated radio spectral index $\alpha$ on the SFR, which is indicated by the weak correlation between them, but the large scatter prevents us from further constraining a well-defined $\alpha$-SFR relation. Future observations in broader bands with comparable quality to the CHANG-ES program are needed to confirm the change of the radio-IR slope with the frequency of the radio continuum emission.

\bigskip
\noindent\textbf{\uppercase{acknowledgements}}
\smallskip\\
\noindent The authors would like to thank Basu A., Heesen V., Calzetti D., and the anonymous referee for very helpful discussions and comments. Li J.-T. acknowledges the financial support from NASA through the grants NNX13AE87G, NNH14ZDA001N, NNX15AM93G, and NNX15AV24G. The National Radio Astronomy Observatory is a facility of the National Science Foundation operated under cooperative agreement by Associated Universities, Inc. The work at Ruhr-University Bochum has been supported by DFG through FOR1048. CJV acknowledges support for this work from the National Science Foundation Graduate Research Fellowship under Grant No. DGE-1144468.

\end{document}